%% file: Indice.tex
\documentclass[11pt]{report}
\usepackage{a4,color,graphics}
\usepackage{graphicx}
\usepackage{amsmath}
\usepackage{amssymb}
\usepackage{amsthm}
\usepackage{amsfonts}
\usepackage{amsxtra}
\usepackage{amsbsy}
\usepackage{amscd}
\usepackage{amsopn}
\usepackage{amstext}
\usepackage[left=1.5in, right=1in, top=1in, bottom=1in, includefoot, headheight=13.6pt]{geometry}
\usepackage{setspace}
\usepackage{subfigure}
\usepackage{booktabs}

\title{\resizebox{!}{1cm}{An equilibrium approach to modelling social interaction}}
\author{\resizebox{!}{0.8cm}{Ignacio Gallo}}
\date{\resizebox{!}{0.8cm}{\today}}

\theoremstyle{definition}

\theoremstyle{plain}
\newtheorem{theorem}{Theorem}

\newtheorem{lemma}{Lemma}

\def\no{\noindent}
\def\be{\begin{equation}}
\def\ee{\end{equation}}
\def\bea{\begin{eqnarray}}
\def\eea{\end{eqnarray}}

\def\<{\langle}
\def\>{\rangle}
\def\~{\tilde}
\def\s{\sigma}
\def\l{\lambda}

\def\a{\alpha}

\def\ee{\varepsilon}

\def\g{\gamma}

\def\o{\omega}
\def\t{\tau}

\def\ds{\displaystyle}

\newcommand{\R}{\mathbb R}
\newcommand{\E}{\Bbb E}

\newcommand{\resettheoremcounters}{%
  \setcounter{theorem}{0}
  \setcounter{prop}{0}
  \setcounter{lemma}{0}
  \setcounter{defi}{0}    }

\newcommand{\ulimit}[1]{\underset{#1}{\longrightarrow}}

\newcommand{\undertilde}[1]{\underset{ \sim }{#1}}

\newcommand{\distr}{\overset{d}{=}}

\def\arctanh{{\rm arctanh \,}}
\def\cosh{{\rm cosh}}
\def\Var{{\rm Var}}

\def\tm{{\tilde m}}

\def\ra{\rightarrow}

\def\de{\partial}

\def\ds{\displaystyle}

\newtheorem{proposition}{Proposition}
\newenvironment{proofof}[1]{\no{\it Proof of #1:}}{\vskip .5cm \hfill$\square$\vskip.5cm}

\newcommand{\captionfonts}{\small\itshape}
\makeatletter

\long\def\@makecaption#1#2{%
\vskip\abovecaptionskip

\sbox\@tempboxa{{\captionfonts #1: #2}}%

\ifdim \wd\@tempboxa >\hsize
  {\captionfonts #1: #2\par}
\else
 \hbox to\hsize{\hfil\box\@tempboxa\hfil}%
\fi

\vskip\belowcaptionskip}

\makeatother

 \makeatletter
    \def\thebibliography#1{\chapter*{References\@mkboth
      {REFERENCES}{REFERENCES}}\list
      {[\arabic{enumi}]}{\settowidth\labelwidth{[#1]}\leftmargin\labelwidth
    \advance\leftmargin\labelsep
    \usecounter{enumi}}
    \def\newblock{\hskip .11em plus .33em minus .07em}
    \sloppy\clubpenalty4000\widowpenalty4000
    \sfcode`\.=1000\relax}
    \makeatother

\begin{document}




\input{frontespiziotesi}

\thispagestyle{empty} \cleardoublepage

\vfill

\begin{flushright}
    {\it to Liliana, Ricardo and Federico. \\ to Sara.}
\end{flushright}

\input{abstract}







\renewcommand{\baselinestretch}{1.27} \small\normalsize

\tableofcontents

\newpage

\pagenumbering{arabic} \setcounter{page}{1}

%
%
%
%

	\input{intro}

	\input{capitolo_discrete_choice}

	\input{capitolo_cw}

	\input{capitolo_cw2d}

	\input{capitolo_case_studies}

\input{ackn}

\input{biblio}
\end{document}

%% file: frontespiziotesi.tex
\begin{titlepage}
\begin{center}
{{\Large{\textsc{Alma Mater Studiorum $\cdot$ Universit\`a di
Bologna}}}} \rule[0.1cm]{15.8cm}{0.1mm}
\rule[0.5cm]{15.8cm}{0.6mm}
{\small{\bf FACOLT\`A DI SCIENZE MATEMATICHE, FISICHE E NATURALI\\
DOTTORATO DI RICERCA IN MATEMATICA, XXI CICLO}}
\end{center}

\begin{center}
    MAT 07: Fisica Matematica
\end{center}

\vspace{27mm}
\begin{center}
{\LARGE{\bf An equilibrium approach}}\\
\vspace{3mm}
{\LARGE{\bf to modelling social interaction}}\\
\vspace{10mm} {\large{\bf doctoral thesis}}
\end{center}
\vspace{22mm}
\par
\noindent

\begin{center}
        {\large{\textsc{Presentata da:}\\
        {\bf Ignacio Gallo}}}
\end{center}

\vspace{15mm}

\begin{minipage}[t]{0.47\textwidth}
{\large{\textsc{Coordinatore:}\\
{\bf Alberto Parmeggiani}}}
\end{minipage}
\hfill
\begin{minipage}[t]{0.47\textwidth}\raggedleft
{\large{\textsc{Relatore}:\\
{\bf Pierluigi Contucci}}}
\end{minipage}
\vspace{20mm}

\begin{center}
\rule[0.5cm]{7cm}{0.1mm}

{\large{\bf Esame Finale anno 2009}}
\end{center}
\end{titlepage}

%% file: abstract.tex
\begin{abstract}

The aim of this work is to put forward a statistical mechanics theory of social interaction, generalizing
econometric discrete choice models. After showing the formal equivalence linking econometric multinomial
logit models to equilibrium statical mechanics, a multi-population generalization of the Curie-Weiss model
for ferromagnets is considered as a starting point in developing a model capable of
describing sudden shifts in aggregate human behaviour.

Existence of the thermodynamic limit for the model is shown by an asymptotic sub-additivity method and
factorization of correlation functions is proved almost everywhere. The exact solution of the model
is provided in the thermodynamical limit by finding converging upper and lower bounds for the system's
pressure, and the solution is used to prove an analytic result regarding the number of possible equilibrium
states of a two-population system.

The work stresses the importance of linking regimes predicted by the model to real phenomena, and
to this end it proposes two possible procedures to estimate the model's parameters starting from micro-level
data. These are applied to three case studies based on census type data: though these studies are found to be
ultimately inconclusive on an empirical level, considerations are drawn that encourage further refinements of
the chosen modelling approach.

\end{abstract}

%% file: intro.tex
\resettheoremcounters

\chapter{Introduction}\label{intro}

In recent years there has been an increasing awareness towards the
problem of finding a quantitative way to study the role played by
human interactions in shaping the kind of aggregate behaviour observed at a population
level: reference \cite{critmass} provides a comprehensive
account of how ramified this field of study already is.
There the author reviews efforts made by researchers from areas as diverse
as psychology, economics and physics, to cite a few,
in the pursuit of regularities that may characterize different kinds
of aggregate human behaviour such as urban traffic, market behaviour
and the internet.

The idea of characterizing society as a unitary entity, characterized
by global features not dissimilar from those exhibited by physical or living
systems has accompanied the development of philosophical thought since
its very beginning, and one must look no further than Plato's {\it Republic}
to find an early example of such a view. The proposal that mathematics might
play a crucial role in pursuing such an idea, on the other hand, dates back
at least to Thomas Hobbes's {\it Leviathan}, where an attempt is made to
draw analogies between the laws describing mechanics, and features of society
as a whole. Hobbes's work gives an inspiring outlook on the ways in which
modern science might contribute to practical human affairs from an organizational
point of view, as well as technological.

In later centuries, nevertheless, quantitative science has grown aware of the fact
that, though a holistic view such as Hobbes's plays an important motivational role
in the development of new scientific enterprises, it is only by reducing a problem
to its simplest components that success is attained by empirical studies.
One of the interesting sub-problems singled out by the modern approach
is that of characterizing the behaviour of a large groups of people, when each individual
is faced with a choice among a finite set of alternatives, and a set of motives driving
the choice can be identified. Such motives might be given by the person's personal
preferences, as well as by the way he interacts with other people. My thesis
aims to contribute to the research effort which is currently analysing the
role played by social interaction in the human decision making process just described.

As early as in the nineteen-seventies the dramatic
consequences of including interaction between peers into a mathematical model
of choice comprising large groups of people
have been recognized independently by the physical \cite{follmer},
economical \cite{schelling} and social science \cite{granovetter}
communities. The conclusion reached by all these studies is that
mathematical models have the potential to describe several features
of social behaviour, among which the sudden and dramatic
shifts often observed in society trends \cite{kuran}, and that
these are unavoidably linked to the way individual people
influence each other when deciding how to behave.

The possibility of using such models as a tool of empirical
investigation, however, is not found in the scientific literature
until the beginning of the present decade \cite{durlauf}: the
reason is to be found  in the intrinsic difficulty
of establishing a methodology of systematic measurement for social
features. Confidence
that such an aim might be an achievable one has been boosted by
the wide consensus gained by econometrics following the Nobel
prize awarded in $2000$ to economist Daniel Mcfadden for his work
on probabilistic models of discrete choice, and by the increasing
interest of policy makers for tools enabling them to cope with the
global dimension of today's society \cite{halpern, gacocoga}.

This has led very recently to a number of studies confronting directly the challenge of quantitatively measuring
social interaction for {\it bottom-up} models, that is, models deriving macroscopic phenomena from assumptions
about human behaviour at an individual level \cite{bouchaud1, salganik, bouchaud2, soet}.

These works show an interesting interplay of methods coming from
econometrics \cite{greene}, statistical physics \cite{galam} and
game theory \cite{nesh}, which reveals a substantial overlap in
the basic assumptions driving these three disciplines. It must also
be noted that all of these studies rely on a simplifying assumption
which considers interaction working on a global uniform scale,
that is on a {\it mean field} approach. This is due to the inability, stated in \cite{watts}, of
existing methods to measure social network topological structure in any detail.
It is expected that it is only matter of time before technology allows to overcome this difficulty: in the meanwhile,
one of the roles
of today's empirical studies is to assess how much information can be derived from the existing kind of data
such as that coming from surveys, polls and censuses.

This thesis considers a mean field model that highlights the possibility of using the methods of discrete choice
econometrics to apply a statistical mechanical generalization of the model introduced in \cite{durlauf}.
The approach is mainly that of mathematical-physics: this means that the main aim shall be to establish the mathematical
properties of the proposed model, such as the existence of the thermodynamical limit, its factorization properties, and its solution,
in a rigorous way: it is hoped that this might be used as a good building block for later more refined theories.
Furthermore, since maybe the most problematic point of a mathematical study of society lies in the feasibility of measuring
the relevant quantities starting from real data, two estimation procedures are put forward: one tries to mimic the econometrics
approach, while the other stems directly from equilibrium statistical mechanics, by stressing the role played by fluctuations
of main observable quantities. These procedures are applied to some simple case studies.

The thesis is therefore organised as follows: the first chapter reviews the theory of Multinomial Logit discrete choice models. These
models are based on a probabilistic approach to the psychology of choice \cite{luce}, which is chosen here as the modelling
approach to human decision making. In this chapter we focus on the mathematical form of Multinomial Logit, and in particular
on its equivalence to the statistical mechanics of non-interacting particles. In the second chapter we consider the
Curie-Weiss model, of which we provide a treatment recently developed in the wider study of mean field spin glasses \cite{guerrarev},
which allows to give elegant rigorous proofs of the model's properties. In chapter three we generalise results from chapter
two for a system partitioned into an arbitrary number of components. Since such a model corresponds to the generalization
of discrete choice first considered in \cite{durlauf}, which includes the effect of peer pressure into the process decision
making, it provides a potential tool for the study of social interaction: chapter four shows an application of this to
three simple case studies.

%% file: capitolo_discrete_choice.tex
\resettheoremcounters

\chapter{Discrete choice models}\label{ds}

In this chapter we describe the general theory of discrete choice models. These
are econometric models that were first applied to the study of demand in
transportation systems in the nineteen-seventies \cite{benakiva}. When people
travel they can
choose the mode of transportation between a set of distinct alternatives, such as
train or automobile, and the basic tenet of these models is that such a {\em discrete
choice} can be described by a probability distribution, and that proposals for
the form of such distribution can be derived from principles established at the
level of individuals. As we shall see this modus operandi is one familiar to
statistical mechanics, and corresponds to what is commonly known as a {\it bottom-up}
strategy in finance.

After describing the general scope of discrete choice analysis, in
section \ref{ML} we describe precisely the mathematical structure of one of the most widely used
discrete choice models, the Multinomial Logit model.
Here we shall see how the probability distribution describing people's choices arises from the assumption
that individual act trying to maximize the benefit coming from that choice, which is
the common setting of neoclassical economics.
Discrete choice models, in general, ignore the effect of social interaction,
but we shall see in subsection \ref{SM} that the Multinomial Logit can be rephrased precisely
as a statistical mechanical model, which gives an ideal starting point
for extending such a model of behaviour to a context including interaction, to be considered
in later chapters.

Due to his development of the theory of the Multinomial Logit model
economist Daniel McFadden was awarded the Nobel Prize in Economics in 2000 \cite{mcfadden}, for
bringing economics closer to quantitative scientific measurement.
The purpose of discrete choice theory is to describe people's behaviour: it is an
econometric technique to infer people's preferences from empirical data. In discrete
choice theory the decision-maker is assumed to make choices that maximise his/her
own benefit. Their `benefit' is described by a mathematical formula, a {\it utility function},
which is derived from data collected in surveys. This utility function includes rational
preferences, but also accounts for elements that deviate from rational behaviour.

\begin{figure}
			    \centering
			    \includegraphics[width=11 cm]{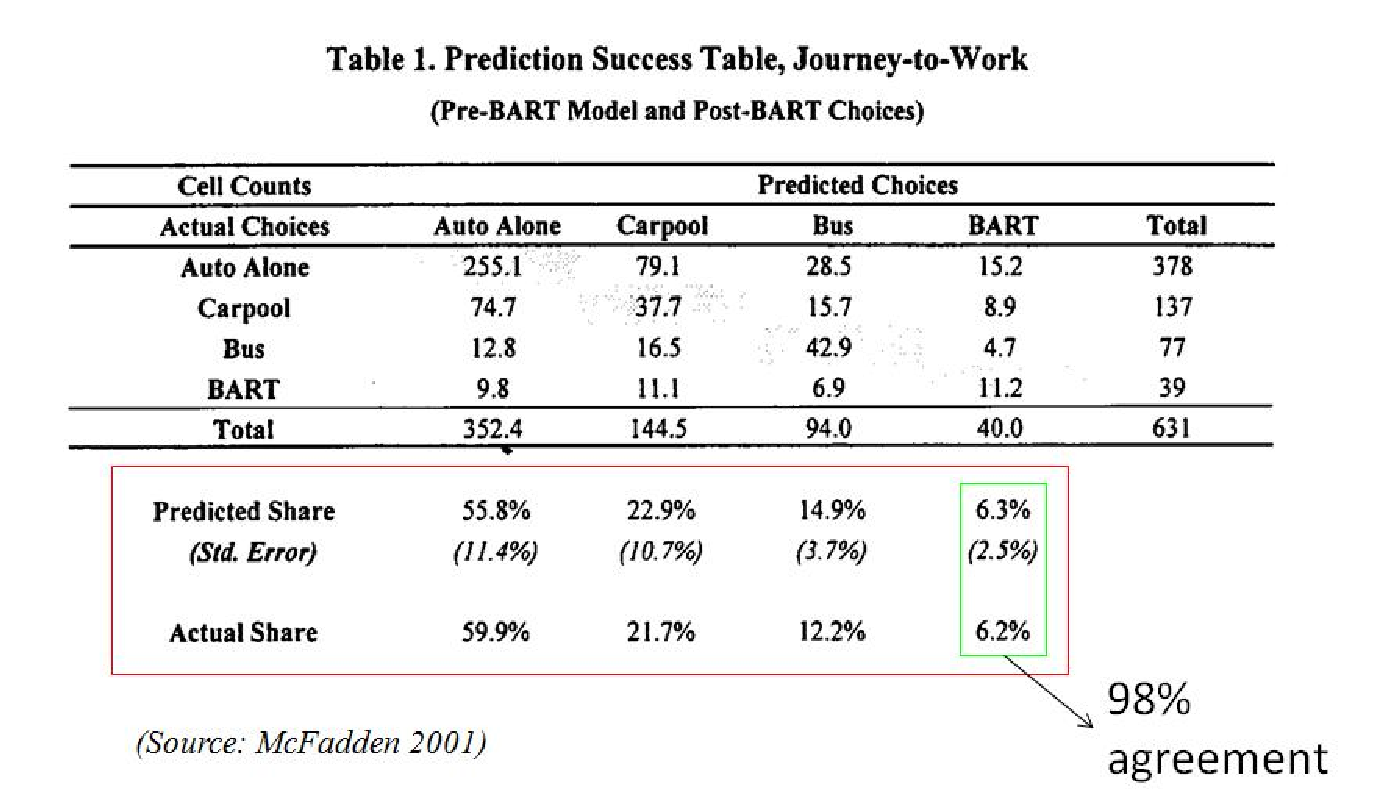}
			    \caption{Discrete choice predictions against actual use of
            travel modes in San Francisco, 1975 (source: McFadden 2001)}\label{mcfadden}
		\end{figure}

Though discrete choice models do not account for `peer pressure'or `herding
effects', it is nonetheless a fact that the standard performance of discrete choice models is close to
optimal for the analysis of many phenomena where peer influence is perhaps not a major
factor in an individual's decision: Figure \ref{mcfadden} shows an example of this. The table (taken from \cite{mcfadden})
compares predictions and actual data concerning use of travel modes, before and after the introduction
of new rail transport system called BART in San Francisco, 1975. We see a remarkable agreement
between the predicted share of people using BART (6.3$\%$),
and the actual measured figure after the introduction of the service (6.2$\%$).

\section{General theory}
In discrete choice each decision process is described mathematically by a {\em utility function}, which each
individual seeks to maximize. The principle of utility maximization is one which lies at the heart of
neoclassical economics: this has often been critised as too simplistic an assumption for complex
human behaviour, and this
criticism has been supported by the poor performance of quantitative models arising from such an assumption.
It must be noted however, that if we wish to attain a quantitative description of human behaviour
at all, we must do so by considering a description which is analytically treatable.
There exist of course alternatives approaches (e.g. agent-based modeling), but since
this field of research is still in its youth, it pays to consider possible improvements of utility maximisation
before abandoning it altogether. This is indeed the view taken by discrete choice, which sees people
as rational utility maximizers, but also takes into account a certain degree of irrationality, which is modeled
through a random contribution to the utility function.

As an example, a binary choice could be to either cycle to work or to
catch a bus. The utility function for choosing the bus may be written as:
\begin{equation}\label{util}
    U=V + \varepsilon
\end{equation}
where V, the deterministic part of the utility, could be symbolically parametrised as follows
\begin{equation}\label{V}
    V=\sum_a \l_a x_a + \sum_a \a_a y_a
\end{equation}

The variables $x_a$ are assumed to be attributes regarding the choice alternatives themselves. For example,
the bus fare or the journey time. On the other hand, the $y_a$ may socio-economic
variables that define the decision-maker, for example their age, gender or income.
It is this latter set of parameters that allows us to zoom in on specific geographical
areas or socio-economic groups.
The $\l_a$ and  $\a_a$ are parameters that need to be estimated empirically, through
survey data, for instance. The key property of these parameters is that they quantify
the relative importance of any given attribute in a person's decision: the larger its value,
the more this will affect a person's choice. For example, we may find that certain people
are more affected by the journey time than the bus fare; therefore changing the fare may
not influence their behaviour significantly. The next section will explain how the value
of these parameters is estimated from empirical data.
It is an observed fact \cite{lucesuppes, ariely} that choices are not always
perfectly rational. For example, someone who usually goes to work by bus may one day
decide to cycle instead. This may be because it was a nice sunny day, or for no evident
reason. This unpredictable component of people's choices is accounted for by the random term $\varepsilon$.
The distribution of $\varepsilon$ may be assumed to be of different forms, giving rise to
different possible models: if, for instance,  $\varepsilon$ is assumed to be normal, the
resulting model is called a {\it probit} model, and it doesn't admit a closed form solution.
Discrete choice analysis assumes  $\varepsilon$ to be extreme-value distributed, and the
resulting model is called a logit model \cite{benakiva}. In practice this is very
convenient as it does not impose any significant restrictions on the model but simplifies it
considerably from a practical point of view. In particular, it allows us to obtain a closed
form solution for the probability of choosing a particular alternative, say catching a bus
rather than cycling to work :
\begin{equation}\label{probds}
P=\frac{e^V}{1+e^V},
\end{equation}
(see section \ref{ML} for the derivation).

In words, this describes the rational preferences of the decision maker.
As will be explained later on, (\ref{probds}) is analogous to the equation describing the
equilibrium state of a perfect gas of heterogeneous magnetic particles (a {\it Langevin paramagnet}):
just like gas particles
react to external forces differently depending, for instance, on their mass and charge,
discrete choice describes individuals as experiencing heterogeneous influences in their
decision-making, according to their own socio-economic attributes, such as gender and wealth.
A question arises spontaneously: do people and gases behave in the same way? The answer to such a
controversial question is that in
some circumstances they might. Models are idealisations of reality, and equation (\ref{probds})
is telling us that the same equation may describe idealised aspects of both human and gas
behaviour; in particular, how individual behaviour relates to macroscopic or societal
variables. These issues go beyond the scope of this thesis, but it is important to note that
(\ref{probds})
offers a mathematical and intuitive link between econometrics and statistical mechanics. The importance
of this `lucky coincidence' cannot be
overstated, and some of the implications will be discussed later on in more detail.

\section{Empirical estimation}

Discrete choice may be seen as a purely empirical model. In order to specify the actual
functional form associated with
a specific group of people facing a specific choice, empirical data is needed. The actual
utility function is then specified by estimating the numerical values of the parameters $\l_a$
and  $\a_a$ which appear in our definition of $V$ given by (\ref{V}), thus establish the
choice probabilities (\ref{probds}). As mentioned earlier,
these parameters quantify the relative importance of the
attribute variables $x_a$ and $y_a$. For example, costs are always associated with negative
parameters: this means that the higher the price of an alternative, the less likely people
will be to choose it. This makes intuitive sense: what discrete choice offers is a
quantification of this effect.
Once the data has been collected, the model parameters may be estimated by standard statistical techniques:
in practice, Maximum Likelihood estimation methods are used most often (see, e.g., \cite{benakiva}
chapter 4). We shall see in further chapters how, though optimal for standard discrete choice models, Maximum likelihood
estimation seems to be unsuitable for phenomena involving interaction due to discontinuities in the
probability structure. As we shall see, a valuable alternative is given by a method put forward by Joseph Berkson \cite{berkson}.

Discrete choice has been used to study people's preferences since the seventies \cite{mcfadden}. Initial applications focused on transport \cite{train, ortuzar}. These models have been used to develop national and regional transport models around the world, including the UK, the Netherlands \cite{fox}, as well as Copenhagen \cite{paag}.
Since then discrete choice has also been applied to a range of social problems, for example healthcare \cite{gerard, ryangerard}, telecommunications \cite{ida} and social care \cite{ryannetten}

\section{The Multinomial logit model}\label{ML}

The binomial logit model which gives the probabilities (\ref{probds}) can be seen as a special case of the Multinomial Logit
model introduced by R. Duncan Luce
in 1959 \cite{luce} when developing a mathematical theory of choice in psychology, and was later given the
utility maximization form which we describe here by Daniel Mcfadden \cite{mcfadden}.

In the following three subsections we shall describe the mathematical structure of a Multinomial Logit model.
In the first subsection we shall first give information about the Gumbel extreme-distribution, which is the distribution
by which the model
describes the random contribution $\ee$ to a person's utility, and is chosen essentially for reasons
of analytical convenience. The second subsection uses the properties of Gumbel distribution in order
to derive the probability structure of the model.
 These two sections are an `executive summary' of all the main things, and they can be found on any
 standard book
 on econometrics \cite{benakiva, greene}.

The third subsection gives the statistical mechanical reformulation of the Multinomial Logit model,
by showing that the same probability structure arises when we compute the pressure of a suitably
chosen Hamiltonian: this leads the way for the extensions of the model that shall be considered
in later chapters.

\subsection{Properties of the Gumbel distribution}

In order to implement the modelling assumption of utility maximization in a quantitative way,
we need a suitable probability distribution for the random term $\ee$.

The Multinomial Logit Model models randomness in choice by a Gumbel distribution, which
has a cumulative distribution function
$$
F(x)=\exp\{- e^{-\mu(x-\eta)} \}, \quad \mu>0,
$$
and probability density function
$$
f(x)=\mu e^{-\mu (x-\eta)} \exp \{ -\mu(x-\eta) \}.
$$

We have that if $\ee \distr {\rm Gumbel(\eta, \mu)}$ then
$$
\E(\ee)= \eta + \frac{\g}{\eta}, \quad \Var(\ee) =\frac{\pi^2}{6 \mu^2},
$$
where $\g$ is the Euler-Mascheroni constant ($\cong 0.577$).

The Gumbel distribution is a type of extreme-value distribution, which means that under suitable
conditions it gives the limit distribution for the value of the extremum of a sequence of i.i.d
random variables, just like the Gaussian distribution does for their average under the central
limit theorem.
In econometrics the Gumbel distribution is used for mainly analytical reasons, since it has
a number of interesting properties, which make it suitable as a modeling tool. As
we shall see in subsection
\ref{SM} the model that one obtains can be readily mapped into
a statistical mechanical model, thus establishing an interesting
link between economics and physics.

The following two properties regard Gumbel variables with equal variance,
and hence equal $\mu$ (see \cite{benakiva}, pag. 104).

\begin{itemize}
    \item[{\bf I.}]  If $\ee' \distr {\rm Gumbel}(\eta_1,\mu)  $ and $\ee'' \distr {\rm Gumbel}(\eta_2,\mu)$ are independent
                random variables, then
                    $\ee=\ee'-\ee''$ is logistically distribute with cumulative distribution
                $$
                F_\ee(x)=\frac{1}{1 + e^{-\mu(\eta_2-\eta_1-x)}},
                $$
                and probability density
             $$
                f_\ee(x)=\frac{\mu e^{-\mu(\eta_2-\eta_1-x)} }{(1+e^{-\mu(\eta_2-\eta_1-x)})^2}.
             $$
    \item[{\bf II.}] If $\ee_i \distr {\rm Gumbel}(\eta_i,\mu)$ for $1 \leqslant i \leqslant k$ are independent then
                $$
                  \max_{i=1..k} \ee_i \distr {\rm Gumbel}\big(\frac{1}{\mu} \ln \sum_{i=1}^k e^{\mu \eta_i},\mu \big )
                $$
\end{itemize}
As we said, the logit is a model which is founded on the assumption that individuals choose their behaviour trying
to maximize a utility, or a ``benefit'' function.
In the next section we shall use Property {\bf II} to handle the probabilistic maximum of the utilities coming from
many different choices, whereas Property {\bf I} shall be used to compare probabilistically the benefits of two
different choices.

\subsection{Econometrics}

We shall now derive the probability distribution for an individual $l$ choosing between $k$ alternatives
$i=1..k$. We have that choice $i$ yields $l$ a utility:
$$
U_i^{(l)}=V_i^{(l)}+\ee_i^{(l)}
$$

We assume that $l$ chooses the alternative with the highest utility. However, since these are random
we can only compute the probability that a particular choice is made:
$$
p_{\, l,i} = P(\, \textrm{`` $l$ chooses $i$ ''} \, )
$$

This is in fact the probability that $U_i^{(l)}$ is bigger than all other utilities, and we can write this
as follows:
\begin{eqnarray*}
p_{\, l,i} = P \big ( U_i^{(l)} \geqslant \max_{j \neq i} U_j^{(l)}  \big)
                                            =  P \big( \, V_i^{(l)} + \ee_i^{(l)} \geqslant \max_{j \neq i} ( \, V_j^{(l)} + \ee_j^{(l)} ) \big )
\end{eqnarray*}

Now define
$$
U^*=\max_{j \neq i} ( \, V_j^{(l)} + \ee_j^{(l)} ).
$$

By property {\bf II} of the Gumbel distribution,
$$
U^* \distr {\rm Gumbel} \big (\frac{1}{\mu} \ln \sum_{j \neq i} e^{\mu V_j^{(l)}}, \mu \big )
$$

So, if
$$
V^*=\frac{1}{\mu} \ln \sum_{j \neq l} e^{\mu V_j^{(l)}},
$$
we have that $U^*=V^*+ \ee^*$ with $\ee^* \distr {\rm Gumbel} (0, \mu)$.

This in turn gives us that
$$
p_{\, l,i} = P( V_i^{(l)} + \ee_i^{/(l)} \geqslant V^* + \ee^* ) =
            P( V_i^{(l)} - V^*  \geqslant \ee^* - \ee_i^{/(l)} ) = \frac{1}{1+e^{\mu(V^*-V_i^{(l)})}}=
$$
by property {\bf I} of the Gumbel distribution, and this can be re-expressed as
$$
p_{\, l,i}=\frac{e^{\mu V_i^{(l)}}}{e^{\mu V_i^{(l)}}+e^{\mu V^*}}=
            \frac{e^{\mu V_i^{(l)}}}{\sum_{j=1}^ke^{\mu V_i^{(l)}}}
$$

According to econometric knowledge $\mu$ is a parameter which cannot be {\it identified} from statistical
data. From a physical perspective, this corresponds to the lack of a well defined temperature: intuitively
this makes sense, since measuring temperature consists in comparing a system of interest with another
system whose state
we assume to know perfectly well. In physics this can be done to a high degree of precision: in social
systems, however, such a concept has yet no clear meaning, and finding one will most certainly require a change
in perspective about what we mean by measuring a quantity.
%
%

As a practical consequence, in this simple model we have that we can let the parameter $\mu$ be incorporated into the
degrees of freedom $V_i^{(l)}$ of the various utilities, and get the choice probabilities in the following
form:
\begin{equation}\label{mlprob}
p_{\; l \! , \, i}= \frac{e^{V_i^{(l)}}}{\sum_{j=1}^ke^{V_i^{(l)}}}
\end{equation}

\subsection{Statistical mechanics}\label{SM}

As we have seen, the Multinomial logit model follows a {\it utility-maximization} approach,
in that it assumes that each person behaves as to optimize his/her own benefit. From
a statistical-mechanical perspective, this amounts to the community of people trying
to identify its {\it ground state}, where some definition of self-perceived
well-being, the utility, takes the role traditionally played by energy.

If there were an exact value of the utility corresponding to each behaviour,
a system characterized by such maximizing principle for the ground state would
identify {\it microcanonical ensemble} in a equilibrium statistical mechanics.
This in amounts to stating that the energy of the system has an exact value,
as opposed to being a random variable.

However, since the Multinomial logit defines utility itself as a Gumbel random
variable in order to try and capture both the predictable and unpredictable components
of human decisions,
 its ``ground state'' turns out to be a ``noisy'' object. Statistical mechanics
models this situation by defining a so-called {\it canonical ensemble}, where all
possible values of the energy are considered, each with a probability given by a {\it Gibbs distribution},
which weights energetically favourable states more than unfavourable ones.
We will now see how the Gibbs distribution leads to a model which is formally equivalent
to the Multinomial logit arising from the Gumbel distribution.

Assume that we have a population of $N$ people, each of whom makes a choice
$$
 \s^{(l)} = \mathbf{e_l}
$$
where vectors $\mathbf{e_i}$ form the $k$-dimensional canonical basis
$$
\mathbf{e_1}=(1,0,.. \, ,0), \quad \mathbf{e_2}=(0,1, .. \, ,0), \quad \textrm{etc}.
$$

We have then that a particular state of this system can be described by the following
set:
$$
\undertilde{\s} = \{ \s^{(1)},...,\s^{(N)} \}
$$

Now define $v^{(l)}$ as a $k$-dimensional vector giving the utilities of the various
choices for individual $l$:
$$
v^{(l)} = (V_1^{(l)}, .. \, , V_k^{(l)}).
$$

We have that $V_i^{(l)}$, which is the deterministic part of the utility considered in the
last section, changes from person to person, and that it can be parametrised by a person's
social attributes, for instance. For the moment, however, we just consider them
as different numbers, since the exact parametrization doesn't change the nature of
the probability structure.

If we now denote by $v^{(l)} \cdot \, \s^{(l)}$ the scalar product between the two vectors,
we may express the energy (also called {\it Hamiltonian}) for the Multinomial Logit Model as follows:
$$
H_N( \, \undertilde{\s} \, )= - \sum_{l=1}^N v^{(l)} \cdot \, \s^{(l)}.
$$

Intuitively, a Hamiltonian model is one where the defines a model where the
favoured states $\undertilde{\s}$ are the ones which make the quantity $H_N$
small, which due to the minus sign, correspond to people choosing as to
maximise their utility. Most of the information contained in an equilibrium
statistical mechanical model can be derived from its pressure, which is defined as
$$
P_N  =  \ln \sum_{\undertilde{\s}} e^{ -H_N(\, \undertilde{\s} \,)},
$$
which acts as a moment generating function for the Gibbs distribution
$$
p(\undertilde{\s})=
\frac{e^{ -H_N(\, \undertilde{\s} \,)}}{\sum_{\undertilde{\s}'} e^{ -H_N(\, \undertilde{\s}' \,)}},
$$
and can recover many of the features of the model, among which the probabilities $p_{\, l,i}$,
as derivatives of $P_N$ with respect to suitable parameters.

This distribution is chosen in physics since it is the one which maximises the system's
entropy at a given temperature, which in turn just means that it is the most likely
distribution to expect for a system which is at equilibrium. This is not to say that
using such a model corresponds to accepting that society is at equilibrium, but
rather to believing that some features of society might have small enough
variations for a period of time long enough to allow a quantitative study. As
pointed out in a later chapter, this belief has at least some quantitative backing
if one considers the remarkable findings made by \'Emile Durkheim as early as at the
end of $19^{th}$ century \cite{durkheim}.

We will now show that this model is equivalent to the Multinomial Logit by computing its
pressure explicitly and finding its derivatives. Indeed, since the model doesn't include
interaction this is a task that can be done easily for a finite $N$:
\begin{eqnarray*}
    P_N  &=&  \ln \sum_{\undertilde{\s}} e^{ -H_N(\, \undertilde{\s} \,)}=
    \ln \sum_{\undertilde{\s}} \exp \big\{ \sum_{l=1}^N v^{(l)} \cdot \, \s^{(l)} \big\}=
\\
        &=& \ln \sum_{ \s^{(1)} } \exp \big\{  v^{(1)} \cdot \, \s^{(1)} \big\} ...
                 \sum_{ \s^{(N)} }  \exp \big\{  v^{(N)} \cdot \, \s^{(N)} \big\}=
\\
        &=& \ln \prod_{l=1}^N \sum_{i=1}^k \exp \{ V_i^{(l)} \}
        = \sum_{l=1}^N \ln \sum_{i=1}^k \exp \{ V_i^{(l)} \}.
\end{eqnarray*}

Once we have the pressure $P_N$ it's easy to find the probability $p_{i,l}$ that person $l$ chooses
alternative $k$, just by computing the derivative of $P_N$ with respect to utility $V^{(l)}_i$:
$$
p_{i,l}=P(\textrm{``$l$ chooses $i$ ''}) = \frac{\de P_N}{\de V^{(l)}_i} = \frac{e^{V^{(l)}_i}}{\sum_{j=1}^k e^{V_j^{(l)}}},
$$
which is the same as (\ref{mlprob}).

This shows how the utility maximization principle is equivalent to a Hamiltonian model, whenever the
random part of the utility is Gumbel distributed. There is a simple interpretation for this statistical
mechanical model: it is a gas of $N$ magnetic particles, each of which has $k$ states, and the energy
of these states depend on the corresponding value of the utility $V^{(l)}_k$, which therefore bears a
close analogy to a magnetic field acting on the particle.

This model may seem completely uninteresting, since it is in no essential way different from a Langevin
paramagnet. What is interesting, however, is how such a familiar, if trivial, model has arisen independently
in the field of economics, and there are a few simple points to be made that can emphasize the change in
perspective.

First, we see how for this model it makes sense to consider the pressure $P_N$ as an extensive quantity.
This is due to the fact that these models are applied to samples of data that yield information about
each single individual, rather than be applied to extremely large ensembles of particles that we regard
as identical, and of which we measure average quantities.
Second, the availability of data about individuals ({\it microeconomic data}) allows us to
define the vector $v^{(l)}$ which assigns a benefit value to each of the alternative that individual
$l$ has.

The main goal of an econometric model of this kind is then to find the parametrization
for $v^{(l)}$ in terms of observable socio-economic features which fits micro data in an optimal way.
The main goal of statistical mechanics is, on the other hand, to find a microscopic theory capable
of generating laws that are observed consistently over a large number of experiments and measured
with extreme precision at a macroscopic level. Since the numbers available for microeconomic data
are not as high as the number of particles in a physical systems, but these that are more detailed
at the level of individuals, the goal of a model of social behaviour could be seen as an interesting mixture
of the above.

\section{The role of statistical mechanics}\label{role}

We have see how discrete choice can be given a statistical mechanical description: in this
section we consider why this is of interest to modeling social phenomena.

A key limitation of discrete choice theory is that it does not formally account for social
interactions and imitation. In discrete choice each individual's decisions are based on purely
personal preferences, and are not affected by other people's choices. However, there is a great
deal of theoretical and empirical evidence to suggest that an individual's behaviour, attitude,
identity and social decisions are influenced by that of others through vicarious experience or
social influence, persuasions and sanctioning \cite{akerlof, bandura}. These theories specifically
relate to the interpersonal social environment including social networks, social support, role
models and mentoring. The key insight of these theories is that individual behaviours and
decisions are affected by their relationships with those around them - e.g. their parents
or their peers.

Mathematical models that take into account social influence have been considered by social
psychology since the '70s (see \cite{scheinkman} for a short review). In particular, influential
works by Schelling \cite{schelling} and Granovetter \cite{granovetter} have shown how models where individuals take
into account the mean behaviour of others are capable of reproducing, at least qualitatively,
the dramatic opinion shifts observed in real life (for example in financial bubbles or during
street riots). In other words, they observed that the interaction built into their models was
unavoidably linked to the appearance of structural changes on a phenomenological level in the
models themselves.
	 \begin{figure}
			    \centering
			    \includegraphics[width=11 cm]{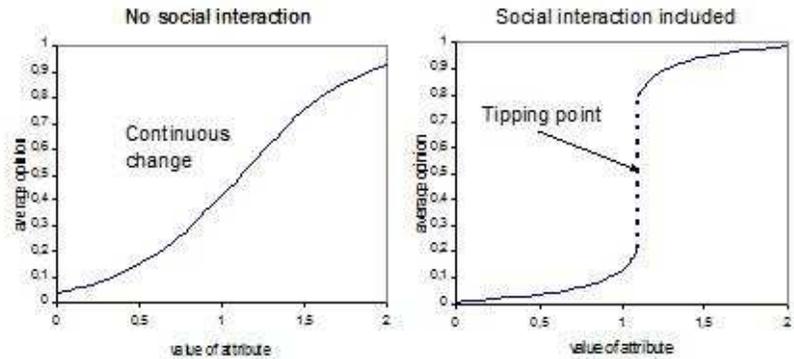}
			    \caption{The diagram illustrates how the inclusion of social interactions (right) leads to the existence tipping points. By contrast models that do not account for social interactions cannot account for the tipping points.  }\label{beforeafter}
		\end{figure}

Figure \ref{beforeafter} compares the typical dependence of average choice with respect to an
attribute parameter, such as cost, in discrete choice analysis (left), where the dependence is
always a continuous one, with the typical behaviour of an interaction model of Schelling or
Granovetter kind (right), where small changes in the attributes can lead to a drastic jump in
the average choice, reflecting structural changes such as the disappearing of equilibria in the
social context.

The research course initiated by Schelling was eventually linked to the parallel development
of the discrete choice analysis framework at the end of the '90s, when Brock and Durlauf \cite{durlauf}
suggested a direct econometric implementation of the models considered by social psychology. In order
to accomplish this, Brock and Durlauf had to delve into the implications of a model where an
individual takes into account the behaviour of others when making a discrete choice: this could
only be done by considering a new utility function which depended on the choices of all other people.

This new utility function was built by starting from the assumptions of discrete choice analysis.
The utility function reflects what an individual considers desirable: if we hold (see, e.g., \cite{bond})
that people consider desirable to conform to people they interact with, we have that, as a consequence,
an individual's utility increases when he agrees with other people.

Symbolically, we can say that when an individual i makes a choice, his utility for that choice
increases by an amount $J_{ij}$ when another individual $j$ agrees with him, thus defining a set of interaction
parameters $J_{ij}$ for all couples of individuals. The new utility function for individual $i$ hence takes
the following form:
\begin{equation}\label{util2}
    U_i=\sum_j J_{ij} \t_j + \sum_a \l_a x_a^{(i)} + \sum_a \a_a y_a^{(i)} + \varepsilon,
\end{equation}
where the sum $\sum_j$ ranges over all individuals, and the symbol $\t_j$  is equal to $1$ if $j$ agrees with $i$,
and $0$ otherwise.

Analysing the general case of such a model is a daunting task, since the choice of
another individual $j$ is itself a random variable, which in turn correlates the choices of all individuals.
This problem, however, has been considered by statistical mechanics since the end of the $19^{th}$ century,
throughout the twentieth century, until the present day. Indeed, the first success of statistical mechanics
was to give a microscopic explanation of the laws governing perfect gases, and this was achieved thanks to a
formalism which is strictly equivalent to the one obtained by discrete choice analysis in (\ref{prob}).

The interest of statistical mechanics eventually shifted to problems concerning interaction between particles,
and as daunting as the problem described by (\ref{util2}) may be, statistical physics has been able to
identify some restrictions on models of this kind to make them tractable while retaining great descriptive
power as shown, e.g., in the work of Pierre Weiss \cite{weiss} regarding the behaviour of magnets.

The simplest way devised by physics to deal with such a problem is called a {\it mean field} assumption,
where interactions are assumed to be of a uniform and global kind. This leads to manageable closed
form solution and a model that is consistent with the models of Schelling and Granovetter.
Moreover, this assumption is also shown by Brock and Durlauf to be closely linked to the assumption
of rational expectations from
economic theory, which assumes that the observed behaviour of an individual must be consistent
with his belief about the opinion of others.

By assuming {\it mean field} or {\it rational expectations} we can rewrite (\ref{util2}) in the
tamer form
\begin{equation}\label{util3}
    U_i=Jm + \sum_a \l_a x_a^{(i)} + \sum_a \a_a y_a^{(i)} + \varepsilon,
\end{equation}
where $m$ is the average opinion of a given individual, and this average value is coupled
to the model parameters by a closed form formula.

If we now define $V_i$  to be the deterministic part of the utility, similarly as before,
$$
V_i=Jm + \sum_a \l_a x_a^{(i)} + \sum_a \a_a y_a^{(i)},
$$
we have that the functional form of the choice probability, given by \ref{prob},
    \begin{equation}\label{prob2}
        P_i=\frac{e^{V_i}}{1+e^{V_i}},
    \end{equation}
remains unchanged, allowing the empirical framework of discrete choice analysis to be used to test
the theory against real data. This sets the problem as one of heterogeneous interacting particles,
and we shall see in the next two chapters how such a mean-field model, just like the standard
Multinomial Logit, can be given a Hamiltonian statistical mechanical form, and solved
in a completely rigorous way using elementary mathematics, via methods recently developed
in the context of spin glasses \cite{guerrarev}.

Though the mean field assumption might be seen as a crude approximation, since it considers
a uniform and fixed kind of interaction, one should bear in mind that statistical physics has
built throughout the twentieth century the expertise needed to consider a wide range of forms
for the interaction parameters $J_{ij}$, of both deterministic and random nature, so that a
partial success in the application of mean field theory might be enhanced by browsing through
a rich variety of well developed, though analytically more demanding, theories.

Nevertheless, an empirical attempt to assess the actual descriptive and predictive power of
such models has not been carried out to date: the natural course for such a study would be to
start by empirically testing the mean field picture, as it was done for discrete choice in the
seventies (see Figure 1), and to proceed by enhancing it with the help available from the
econometrics, social science, and statistical physics communities. Two recent examples
of empirical studies of mean-field models can be found in \cite{soet} and \cite{gabaco}.

%% file: capitolo_cw.tex
\resettheoremcounters

\chapter{The Curie-Weiss model}\label{cw}

The Curie-Weiss model was first introduced in 1907 by Pierre Weiss \cite{weiss} as a proposal
for a phenomenological model
capable of explaining the experimental observations carried out by Pierre Curie in 1895 \cite{curie},
concerning the dependance
on temperature of the magnetic nature for metals such as iron, nickel, and magnetite.

Iron and nickel are materials capable of retaining a degree of magnetization, which we call {\it spontaneous magnetization},
after having been exposed to a magnetic field: such materials are said to be {\it ferromagnetic}, from the Latin name for iron.
However, it had been known since the day of Faraday (\cite{curie}, pag. 1) that these materials tend to lose their ability to retain
magnetization as their temperature increases.

	 \begin{figure}
			    \centering
			    \includegraphics[width=11 cm]{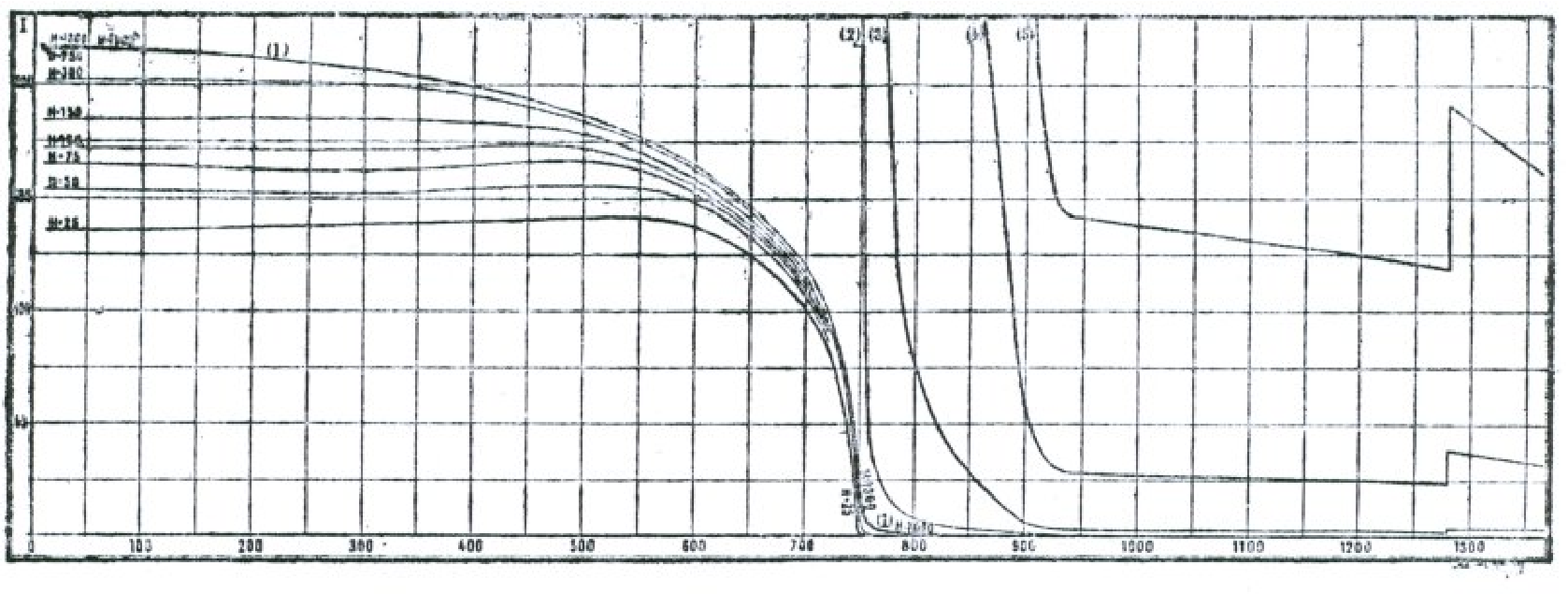}
			    \caption{Pierre Curie's measurements in 1895}\label{curie}
		\end{figure}

Pierre Curie's experiments showed not only that the loss of the ferromagnetic property indeed occurs, but
also observed that it occurs in a very peculiar fashion. For each of the materials he considered, he found
a definite temperature at which
spontaneous magnetization vanishes abruptly, giving rise to an irregular point in the graph plotting
spontaneous magnetization versus temperature (see Figure \ref{curie}):
we now call this temperature the {\it Curie temperature}
for the given material.

Weiss's model arises from physical considerations about the nature of magnetic interactions between atoms: he claims
that single atoms must experience, as well as the external field, a sum of all the fields produced by all
the other particles inside the material. He calls this field a ``molecular field'' ({\it champ moleculaire}), and by
adding a term corresponding to this field inside the balance equation derived by Paul Langevin to describe
{\it paramagnetic} materials (that is, magnetic materials that do not retain magnetition after exposure to a field),
he formulates a balance equation for ferromagnetic materials.

	 \begin{figure}
			    \centering
			    \includegraphics[width=8 cm]{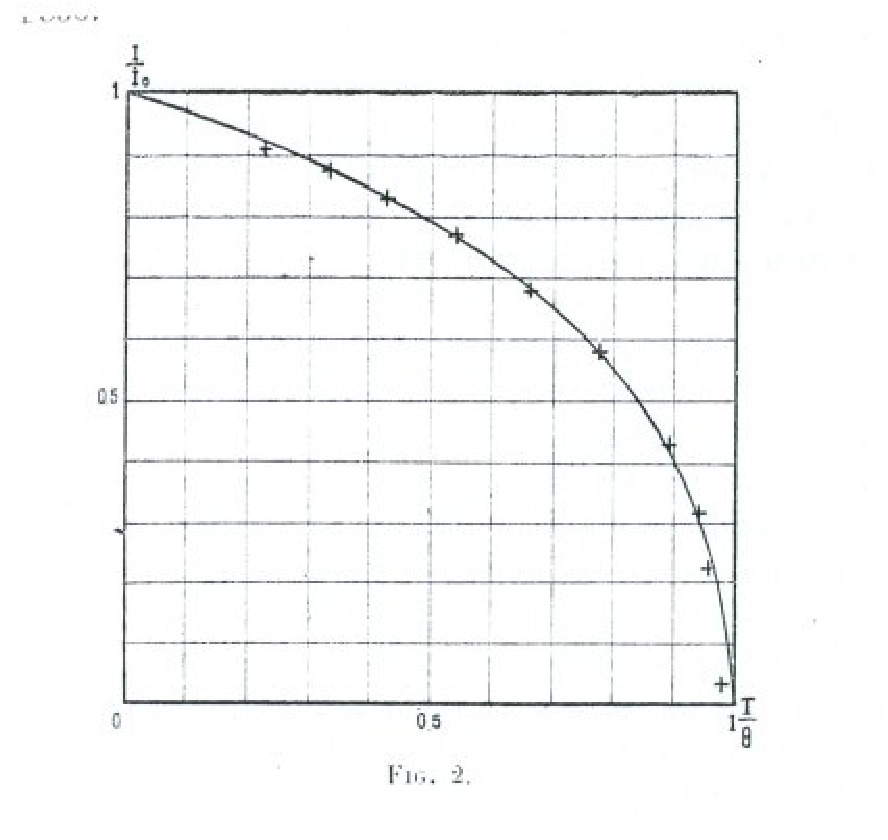}
			    \caption{Pierre Weiss's measurements (crosses) fitted against his theoretical prediction (line) in 1907:
			                   the graph shows the dependance of spontaneous magnetization on temperature for magnetite}
			                   \label{weiss}
		\end{figure}

In his 1907 paper Weiss shows that the theoretical predictions of his model show remarkable agreement with
physical reality by fitting them against measurements, carried on by himself, on a ellipsoid made of magnetite
 (Figure \ref{weiss}).

Today we know that the Curie-Weiss is not completely accurate: indeed, it is well known that some physically measurable
quantities for ferromagnetic materials, called {\it critical exponents}, are not predicted correctly by it (see
\cite{huang}, pag. 425). The subsequent study of more detailed models, such as the Ising model, has brought
 to light
the reason for such a mismatch: when rewritten in the language of modern statistical mechanics, the model of
Curie-Weiss readily shows to be equivalent to one where all particles are interacting with each other.
This turns out to be too strong
an assumption for a system where all particles sit next to each other geometrically and which interact,
according to quantum
mechanics, up to a very short range. On the other hand though, the Ising model, which still makes use of all of Weiss's other
simplifying assumptions about interaction between particles, manages to predict critical exponents correctly, just by
assuming that particles only interact with their nearest neighbours on a regular lattice, though, from a
mathematical point of view, this modification implies a drastic reduction
of the symmetry of the problem, which has so far proved to be analitically untreatable in more
than two dimensions (see \cite{huang} pag. 341).

All objections standing, it is nevertheless worth remembering that  the degree of agreement between theory and reality
for the Curie-Weiss model is truly remarkable
given the simplicity of the model. Today, Weiss's ``molecular field'' assumption is called a {\it mean field}
assumption, and scientific wisdom tells that this assumption is of great value in exploring the phase structure of a system
so that, when faced with a new situation, one would try mean field first (\cite{huang}, pag. 423).

\section{The model}

As a modern statistical mechanics model, the Curie-Weiss model is defined by its Hamiltonian:
\begin{equation}\label{Ham}
    H(\sigma)=-\sum_{i, j=1}^{N}J_{ij}\sigma_{i}\sigma_{j}-\sum_{i=1}^N h_{i}\sigma_{i} \;.
\end{equation}

We consider Ising spins, $\sigma_i=\pm 1$, subject to a uniform magnetic field $h_i=h$
and to isotropic interactions $J_{i,j}=J/2N$, so that we have.
\begin{equation}\label{Ham}
    H(\sigma)=-\frac{J}{2N}\sum_{i, j=1}^{N}\sigma_{i}\sigma_{j}-h\sum_{i=1}^N\sigma_{i} \;.
\end{equation}

If we now introduce the magnetization of a configuration $\sigma$ as
\begin{displaymath}
m(\sigma) \; = \; \frac{1}{N}\sum_{i =1 }^N\sigma_i
\end{displaymath}
we can rewrite the Hamiltonian per particle as:
\begin{equation}\label{Ham}
	\frac{H(\sigma)}{N}=- \frac{J}{2} m(\s)^2\! -  h m(\s)
\end{equation}

The established statistical mechanics framework defines the equilibrium value of an observable $f(\sigma)$ as the
average with respect to the {\it Gibbs distribution} defined by the Hamiltonian. We call this average the {\it Gibbs state}
for $f(\sigma)$, and write it explicitly as:
\begin{displaymath}
\< \, f \, \> \; = \; \frac{\sum_{ \sigma} f(\sigma) \; e^{\,-H(\sigma)}    }{\sum_\sigma e^{\,-H(\sigma)}} \; .
\end{displaymath}

The main observable for our model is the average value of a spin configuration,
i.e. the {\it magnetization}, $m(\sigma)$, which explicitly reads:
\begin{displaymath}
m(\sigma)=\frac{1}{N}\sum_{i=1}^N \sigma_i.
\end{displaymath}

Our quantity of interest is therefore $\<m\>$: to find it, as well as the moments of many other observables,
statistical mechanics leads us to consider the pressure function:
\begin{displaymath}
p_N \; = \; \frac{1}{N}\log \sum_{\sigma}e^{-H(\sigma)} \; .
\end{displaymath}

It is easy to verify that, once it's been derived exactly, the pressure is capable
of generating the Gibbs state for the magnetization as
$$
\< m \> =  \frac{\partial p_N}{\partial h}.
$$

\section{Existence of the thermodynamic limit}\label{exists}

We show two ways of computing the existence of the thermodynamic limit in the Curie-Weiss model.
The first method follows \cite{barra} in exploiting directly the convexity of the Hamiltonian in order to
prove subadditivity in $N$ for the systems's pressure.

The second method consists in a refinement of the first, and covers models for which the Hamiltonian
is not necessarily convex, such as the two-population model considered in the next chapter. It is important
to point out that a careful application of this method to the Sherrington-Kirkpatrick spin glass model allowed
Guerra \cite{guerra} to prove the twenty-years standing question concerning existence of thermodynamic limit.

\subsection{Existence by convexity of the Hamiltonian}

We consider a system of $N$ spins defined as above. Following \cite{barra} we split the system in
two subsystem of $N_1$ and $N_2$ spins, respectively, with $N_1+N_2=N$. For each of these systems
we define partial magnetizations
$$
m_1(\sigma) \; = \; \frac{1}{N_1}\sum_{i =1 }^{N_1}\sigma_i
\quad \textrm{and} \quad
m_2(\sigma) \; = \; \frac{1}{N_2}\sum_{i =N_1+1 }^N\sigma_i,
$$
which allow us to define partial Hamiltonians
$$
H_{N_1}=-N_1(\frac{J}{2}m_1^2+hm_1)
\quad \textrm{and} \quad
H_{N_2}=-N_2(\frac{J}{2}m_2^2+hm_2).
$$

We have by definition that
\begin{equation}\label{partmagn}
	m=\frac{N_1}{N}m_1+\frac{N_2}{N}m_2
\end{equation}
and since $f(x)=x^2$ is a convex function we also have that
\begin{equation}\label{convmagn}
	m \leqslant \frac{N_1}{N}m_1^2+\frac{N_2}{N}m_2^2.
\end{equation}

We are now ready to prove the following

\begin{proposition}\label{existence1}
There exists a function $p(J, h)$ such that
$$
\lim_{N \rightarrow \infty} p_N= p \; .
$$
\end{proposition}

\begin{proof}
Relations (\ref{partmagn}) and (\ref{convmagn}) imply that
\begin{equation*}
	H_N \leqslant H_{N_1}+H_{N_2}
\end{equation*}
and this in turn gives
\begin{equation*}
	Z_N=\sum_\s e^{-H_N(\s)} \leqslant \sum_\s e^{-H_{N_1}(\s:1..N_1)-H_{N_1}(\s:N_1+1..N_2)}= Z_{N_1}Z_{N_2}
\end{equation*}
where $\s:1..N_1=\{\s_1,..,\s_{N_1}\}$ and $\s:N_1+1..N=\{\s_{N_1+1},..,\s_N\}$.
Hence we have the following inequality
$$
Np_N \leqslant N_1p_{N_1}+N_2p_{N_2}, \quad \textrm{for $N_1+N_2=N$}
$$

This identifies the sequence $\{Np_N\}$ as a {\it subadditive} sequence, for which the following holds
$$
\lim_{N \ra \infty} \frac{Np_N}{N}=\lim_{N \ra \infty} p_N=\inf_{N} p_N.
$$

Hence in order to verify the existence of a finite limit we need to verify that the sequence
$\{p_N\}$ is bounded below, which follows from the boundedness of the intensive quantity
$$
\frac{H(\s)}{N}=-\frac{J}{2}m^2-hm,
$$
for $-1\leqslant m \leqslant 1$. Indeed, if $\frac{H(\s)}{N} \leqslant K$,
$$
p_N = \frac{1}{N} \ln \sum_{\s} e^{-H(\s)} \geqslant \frac{1}{N} \ln 2^N e^{NK} = \ln 2+ K
$$
so the result follows.

\end{proof}

\subsection{Existence by interpolation}

We shall now prove that our model admits a thermodynamic limit by exploiting an existence theorem
provided for mean field models in \cite{bcg}: the result states that the existence of the pressure
per particle for large volumes is guaranteed by a monotonicity condition on the equilibrium
state of the Hamiltonian.
We therefore prove the existence of the thermodynamic limit independently of an exact solution. Such
a line of enquiry is pursued in view of the study of models, that shall possibly involve
random interactions of spin glass or random graph type, and that might or might not come with
an exact expression for the pressure.
\begin{proposition}\label{existence2}
There exists a function $p(J, h)$ such that
$$
\lim_{N \rightarrow \infty} p_N= p \; .
$$
\end{proposition}
\vskip .1 in

\begin{proof}
Theorem 1 in \cite{bcg} states that given a Hamiltonian $H_N$ such that $\frac{H_N}{N}$ is
bounded in $N$, and its associated equilibrium state
$\omega_N$, the model admits a thermodynamic limit whenever the physical condition
\begin{equation}\label{additivity1}
\omega_N (H_N) \geqslant \omega_N (H_{N_1}) + \omega_N (H_{N_2}), \qquad N_1 + N_2= N,
\end{equation}
\no is verified.

For the Curie-Weiss model the condition is easy to verify once we define partial magnetizations
$$
m_1(\sigma) \; = \; \frac{1}{N_1}\sum_{i =1 }^N\sigma_i
\quad \textrm{and} \quad
m_2(\sigma) \; = \; \frac{1}{N_2}\sum_{i =1 }^N\sigma_i.
$$

This gives that
$$
m=\frac{N_1}{N}m_1+\frac{N_2}{N}m_2
$$
so that
\begin{eqnarray*}
	H_N-H_{N_1} - H_{N_2}&=&-N(\frac{J}{2}m^2+hm)+N_1(\frac{J}{2}m_1^2+hm_1)+N_2(\frac{J}{2}m_2^2+hm_2)=
\\						
						&=& -N\frac{J}{2}(m^2 - \frac{N_1}{N} m_1^2- \frac{N_2}{N} m_2^2)
						 	-Nh(m - \frac{N_1}{N} m_1- \frac{N_2}{N} m_2)
\\							
						&=& -N\frac{J}{2}(m^2 - \frac{N_1}{N} m_1^2- \frac{N_2}{N} m_2^2) \geqslant 0
\end{eqnarray*}

The last inequality follows from convexity of the function $f(x)=x^2$, and since it holds for every
configuration $\s$, it also implies (\ref{additivity1}), proving the result.

\end{proof}

\section{Factorization properties}

In this section we shall prove that the correlation functions of our model factorize completely
in the thermodynamic limit, for almost every choice of parameters. This implies that
all the thermodynamic properties of the system can be described by the magnetization.
Indeed, the exact solution
of the model to be derived in the next section comes as an equation of state which,
as expected, turns out to be the same as the balance equation derived by Weiss.

\begin{proposition}\label{magnfact}
$$
\lim_{N \rightarrow \infty} \big ( \omega_N(m^2)-\omega_N(m)^2 \big ) =0
$$
\no for almost every choice of $h$.
\end{proposition}

\begin{proof}
We recall the definition of the Hamiltonian per particle
\begin{equation*}
\frac{H_N(\sigma)}{N}=-\frac{J}{2}m^2\! - h m,
\end{equation*}
and of the pressure per particle
$$
p_N=\frac{1}{N} \ln \sum_{\sigma} e^{-H_N(\sigma)}.
$$

By taking first and second partial derivatives of $p_N$ with respect to $h$ we get
$$
\frac{\partial p_N}{\partial h}=\frac{1}{N} \sum_{\sigma}  N m(\sigma) \frac{e^{-H(\sigma)}}{Z_N}= \omega_N(m),
\qquad
\frac{\partial^2 p_N}{\partial \; h^2}= \omega_N(m^2)-\omega_N(m)^2.
$$

By using these relations we can bound above the integral with respect to $h$ of the fluctuations of $m$ in the Gibbs state:
\begin{eqnarray}\label{average var1}
\Bigg |\int_{h^{(1)}}^{h^{(2)}} ( \omega_N(m^2)-\omega_N(m)^2 ) \; dh \Bigg | & = &
\frac{1}{N} \Bigg | \int_{h^{(1)}}^{h^{(2)}}   \frac{\partial^2 p_N}{\partial h^2}   \; dh \Bigg | =
\frac{1}{N} \Bigg | \frac{\partial p_N}{\partial \; h} \bigg |_{h^{(1)}}^{h^{(2)}} \Bigg | \leqslant
\nonumber  \\
& \leqslant & \frac{1}{N} \big ( \big | \omega_N(m)|_{h^{(2)}} \big | + \big | \omega_N(m)|_{h^{(1)}} \big | \big ) =
O \big (\frac{1}{N} \big ).
\nonumber  \\
\end{eqnarray}

On the other hand we have that
$$
\omega_N(m)=\frac{\partial p_N}{\partial h},
$$
and
$$
\omega_N(m^2)=2\frac{\partial p_N}{\partial J},
$$
so, by convexity of the thermodynamic pressure $\ds p=\lim_{N \rightarrow \infty} p_N$,
both quantities $\ds \frac{\partial p_N}{\partial h}$ and $\ds \frac{\partial p_N}{\partial J}$ have well
defined thermodynamic limits almost everywhere.
This together with (\ref{average var1}) implies that
\begin{equation}\label{ae m}
\lim_{N \rightarrow \infty} ( \omega_N(m^2)-\omega_N(m)^2 ) = 0 \quad \textrm{a.e. in $h$}.
\end{equation}

\end{proof}

The last proposition proves that $m(\s)$ is a {\it self-averaging} quantity, that is, a random quantity
whose fluctuations vanish in the thermodynamic limit. This is indeed a powerful result, which
can be exploited thanks to the following
\begin{proposition}{(Cauchy-Schwartz inequality)}\label{CauSch}
	Let $X$ and $Y$ be two random variables defined on a finite probability
	space such that $P(X_i)=P(Y_i)=p_i$. Then the following holds
	$$
	\E(XY)-\E(X)\E(Y) \leqslant \sqrt{\Var(X)\Var(Y)}
	$$
\end{proposition}

\begin{proof}

Let us define the following quantities:
$$
\E(X)=\sum_i X_i p_i=\mu_X, \quad \Var(X)=\s^2_X
$$
$$
\E(Y)=\sum_i Y_i p_i=\mu_Y, \quad \Var(Y)=\s^2_Y
$$

If we now define rescaled versions of $X$ and $Y$:
$$
\bar X=\frac{X-\mu_X}{\s_X}, \quad \textrm{and} \quad \bar Y=\frac{Y-\mu_Y}{\s_Y},
$$
we get that $\{\bar X_i p_i^{1/2}\}$ and $\{\bar Y_i p_i^{1/2}\}$ are vectors of Euclidean
length equal to 1 (since their lengths are the variances of $\bar X$ and $\bar Y$, which have been
normalized). This implies
\begin{equation}\label{euc length}
| \E(\bar X \bar Y)|=|\sum_{i} \bar X_i \bar Y_i p_i |=|\sum_{i} \bar X_i p_i^{1/2}
											\bar Y_i p_i^{1/2}|
						\leqslant 1
\end{equation}
where the inequality only points out that $\E(\bar X \bar Y)$ is the projection
of a unit vector against another, and therefore that its modulus is less than one.

If we now substitute back $X$ and $Y$ into (\ref{euc length}) we get our result.

\end{proof}

By putting together the self-avering property and the Cauchy-Schwartz inequality we get the
following

\begin{proposition}\label{magnfact}
Given any integer $k$ we have that
$$
\lim_{N \rightarrow \infty} \big ( \omega_N(m^k)-\omega_N(m)^k \big ) =0
$$
\no for almost every choice of $h$.
\end{proposition}
\begin{proof}
Applying the Cauchy-Schartz inequality to $X=m^{k-1}$ and $Y=m$ we get that
\begin{equation}\label{firststep}
	|\o_N(m^{k-1}m)-\o_N(m^{k-1})\o_N(m)| \leqslant \sqrt{\Var_N(m^{k-1})\Var_N(m)}.
\end{equation}

Now self-averaging tells us that $\Var_N(m)$ tends to zero in the limit, and since
$m^{k-1}$ is a bounded quantity, (\ref{firststep}) implies:
$$
	\lim_{N \rightarrow \infty} \big ( \omega_N(m^k)-\omega_N(m)^{k-1}\omega_N(m) \big ) =0
$$
and the rest of the proposition follows by induction on the same argument.

\end{proof}

The last proposition is very important for this model, because the mean-field nature of
the system allows to use the factorization of the magnetization in order to prove factorization
of spin correlation functions, thus characterizing all the thermodynamics of the system.

In the following proposition we shall only prove the factorization of 2-spins: the factorization
of k-spins is done in the same way.

\begin{proposition}\label{spinfact}
$$
\lim_{N \rightarrow \infty} \big ( \omega_N(\sigma_i\sigma_j)-\omega_N(\sigma_i)\omega_N(\sigma_j) \big ) =0
$$
\no for almost every choice of $h$, whenever $\sigma_i$, $\sigma_j$
are distinct spins.
\end{proposition}

\begin{proof}
Now we can use the self-averaging of $m(\s)$
the factorization of correlation functions. This is done by exploiting the translation
invariance of the Gibbs measure on spins, which in turn follows from the mean-field nature of the model:
\begin{eqnarray}\label{s i s j}
\omega_N(m)&=&\omega_N(\frac{1}{N}\sum_{i=1}^{N} \s_i )= \omega_N(\s_1),
\nonumber \\
\omega_N(m^2)&=&\omega_N(\frac{1}{N^2}\sum_{i,j=1}^{N} \s_i \s_j )=
\omega_N(\frac{1}{N^2}\sum_{i\neq j=1}^{N_1} \s_i \s_j )+\omega_N(\frac{1}{N^2}\sum_{i = j=1}^{N} \s_i \s_j ) =
\nonumber \\
&=& \frac{N-1}{N}\omega_N(\s_1 \s_2) + \frac{1}{N}. \nonumber
\\
\end{eqnarray}

We have that (\ref{s i s j}) and (\ref{ae m}) imply
\begin{equation}
\lim_{N \rightarrow \infty} \omega_N(\s_i\s_j)-\omega_N(\s_i)\omega_N(\s_j)=0, \quad \textrm{for a.e. h}
\end{equation}
\no which verifies our statement for all couples of spins $i \neq j$.

\end{proof}

The self-averaging of the magnetization has been proved directly here: this, however, can be
seen as a consequence of the convexity of the pressure. Indeed, the second derivative of any
convex function exists almost everywhere: this is a consequence of the first derivative existing
almost everywhere and being monotonically increasing (se, e.g., \cite{royden}).

Therefore existence almost everywhere of $\frac{\de^2 p}{\de h^2}$ together with the intensivity
property of the magnetization implies trivially that its fluctuations vanish in the thermodynamic limit.
This also implies that, since energy per particle is another intensive quantity which is obtained by
differentiating the pressure with respect $J$, energy per particle is a self-averaging quantity too.

As we can see from Proposition \ref{magnfact} factorization of spins only holds a.e. for $h$, and
indeed it can be proved that factorization
doesn't hold at $h=0$, $J>1$. However, by using the self-averaging of energy-per-particle proved above,
we can similarly obtain a weaker factorization rule which covers this regime:

\begin{proposition}\label{enerfact}
\begin{equation*}
\lim_{N \rightarrow \infty} \omega_N(\s_i\s_j\s_k\s_l)-\omega_N(\s_i\s_j)\omega_N(\s_k\s_l)=0, \quad \textrm{for a.e. $J$}
\end{equation*}
\no for almost every choice of  $J$, whenever $\sigma_i$, $\sigma_j$, $\sigma_k$, $\sigma_l$  are distinct spins.
\end{proposition}

\begin{proof}
The proof follows the same argument of Proposition \ref{magnfact}, and uses the self-averaging of
the energy per particle instead of the self-averaging of the magnetization.

\end{proof}

\section{Solution of the model}

We shall derive upper and lower bounds for the thermodynamic limit of the pressure. The lower bound is obtained through the standard entropic variational principle, while the upper bound
is derived by a decoupling strategy.

\subsection{Upper bound}

In order to find an upper bound for the pressure we shall divide the configuration space into a partition of
microstates of equal magnetization, following \cite{desanctis, guerrarev, guerrarev2}. Since the system
 consists of $N$ spins, its magnetization can take
exactly $N+1$ values, which are the elements of the set
$$
R_{N}=\Big \{-1, -1+\frac {1} {2 N}, \dots ,1 -\frac {1} {2 N}, 1 \Big \}.
$$

Clearly for every $m(\sigma)$ we have that
$$
\sum_{\bar m \in R_{N}} \delta_{m, \bar m}=1,
$$

\no where $\delta_{x,y}$ is a Kronecker delta. Therefore we have that
\begin{eqnarray} \label{deltas1}	
Z_N &= &\sum_\sigma   \exp \big\{ N (\frac{J}{2} m^2+  hm ) \big \} ={}
	 =  \sum_\sigma \sum_{\bar m \in R_{N} }
          \delta_{m, \bar m}
   \exp\big\{  N (\frac{J}{2} m^2+  hm ) \big \}.
\end{eqnarray}

Thanks to the Kronecker delta symbols,
we can substitute $m$ (the average of the spins within a configuration)
with the parameter $\bar m$ (which is not coupled to the spin configurations)
in any convenient fashion.
Therefore we can use the following relation in order to linearize the quadratic term
appearing in the Hamiltonian
\begin{eqnarray*}	
	(m - \bar m)^2 & = & 0,
\end{eqnarray*}
and once we've carried out this substitution into (\ref{deltas1}) we are left with a function
which depends only linearly on $m$:
\begin{eqnarray*} 	
	Z_N &= &\sum_\sigma  \sum_{\bar m \in R_{N} } \delta_{m, \bar m}
				           \exp \big\{ N (\frac{J}{2} (2m \bar m - \bar m^2)+  hm ) \big \}.
\end{eqnarray*}

\no and bounding above the Kronecker deltas by 1 we get

\begin{eqnarray*} 	
	Z_N & \leqslant & \sum_\sigma  \sum_{\bar m \in R_{N} }
				           \exp \big\{ N (\frac{J}{2} (2m \bar m - \bar m^2)+  hm ) \big \}.
\end{eqnarray*}

Since both sums are taken over finitely many terms, it is possible to exchange the order of the
two summation symbols, in order to carry out the sum over the spin configurations, which now
factorizes, thanks to the linearity of the interaction with respect to the $m$s. This way we get:

$$
Z_N \leqslant \sum_{\bar m \in R_{N} } G(\bar m).
$$
where
\begin{eqnarray}\label{g}
	G(\bar m) &=&  \exp\big\{-N\frac{1}{2}J \bar m^2 \big\} \cdot 2^{N}\big( \cosh \big( J \bar m  +  h \big) \big) ^{N} \nonumber\\
\end{eqnarray}

Since the summation is taken over the
range   $R_{N}$ of cardinality $N+1$ we get that the total number of summands
is $N+1$. Therefore
\begin{eqnarray}
Z_N & \leqslant   & (N+1) \sup_{\bar m} G,
 \end{eqnarray}

\no which leads to the following upper bound for $p_N$:
\begin{eqnarray} \label{deltas3}
p_N & = & \frac{1}{N} \ln Z_N  \leqslant  \frac{1}{N} \ln (N+1)\sup_{\bar m} G=
\nonumber \\
	&=&\frac{1}{N} \ln (N+1) +\frac{1}{N} \sup_{\bar m} \ln G \; .
\end{eqnarray}
where the last equality follows from monotonicity of the logarithm.


Now defining the $N$ independent function
\begin{eqnarray*}
	p_{up}(\bar m_1,\bar m_2) =\frac{1}{N} \ln G  =   \ln 2 -\frac{J}{2} \bar m^2  +   \ln \cosh \big( J \bar m +  h \big),
 \end{eqnarray*}

\no and keeping in mind that $\lim_{N \ra \infty} \frac{1}{N}\ln(N+1)=0$, in the thermodynamic limit we get:
\begin{equation}
\limsup_{N \rightarrow \infty} p_N \leqslant  \sup_{\bar m} p_{UP}(\bar m).
\end{equation}

We can summarize the previous computation into the following:

\begin{lemma}
Given a Hamiltonian as defined in (\ref{Ham}), and defining the pressure per particle as $p_N=\frac 1 N \ln Z$,
given parameters $J$ and $h$, the following inequality holds:
$$
				\limsup_{N\rightarrow \infty} p_N \leqslant \sup_{\bar m} p_{up}
$$
\no where
$$
	p_{up}(\bar m) =   \ln 2 -\frac{J}{2} \bar m^2  +   \ln \cosh \big( J \bar m +  h \big),
$$
\no and $\bar m \in [-1,1]$.
\end{lemma}

We shall give two ways of deriving a lower bound for the pressure: indeed, it is important to
keep in mind that having as many bounding tecniques as possible can be a good way of
approaching more refined models.

\subsection{Lower bound by convexity of the Hamiltonian}

\begin{proposition}\label{geom}		
		Given a Hamiltonian as defined in (\ref{Ham}) and its associated pressure per particle
		$p_N=\frac 1 N \ln Z$, the following inequality holds for every $J$, $h$:
		$$
				p_N \geqslant \sup_{-1 \leqslant \bar m \leqslant 1} p_{low}
		$$
		\no where
		$$
			p_{low}(\bar m)= - \frac{J}{2} \bar m^2 + \ln  2 + \ln \cosh( J \bar m + h)	  	$$
\end{proposition}

\begin{proof}
We recall the Hamiltonian per particle written in terms of the configuration's magnetization $m(\s)$:
\begin{equation*}
	\frac{H(\sigma)}{N}=- \frac{J}{2} m^2\! -  h m.
\end{equation*}

Now, given any number $\bar m \in [-1,+1]$, the following holds:
$$
(m-\bar m)^2 \geqslant 0 \ \Rightarrow \  m^2 \geqslant 2 m \bar m- \bar m^2
$$
so that
\begin{eqnarray*}
	p_N &=& \frac{1}{N} \ln Z_N = \frac{1}{N} \ln \sum_\s \exp \{N(\frac{J}{2}m^2 +hm) \} \geqslant
\\	
			&\geqslant & \frac{1}{N} \ln \sum_\s \exp \{ N(Jm \bar m - \frac{J}{2} \bar m^2 +hm)  \} =
\\		
		&=&	\frac{1}{N} \ln \Big ( \exp \{ - \frac{NJ}{2} \bar m^2 \} \sum_\s \exp \{ N(J \bar m m +hm) \} \Big )=
\\
		&=& - \frac{J}{2} \bar m^2 + \frac{1}{N} \ln \Big ( 2^N \cosh( J \bar m + h )^N \Big)=
		  - \frac{J}{2} \bar m^2 + \ln  2 + \ln \cosh( J \bar m + h )
\end{eqnarray*}

This way we get new lower bound which can be expressed as
$$
	p_N \geqslant \sup_{-1 \leqslant \bar m \leqslant 1} p_{low}
$$
where
$$
	p_{low}(\bar m)=- \frac{J}{2} \bar m^2 + \frac{1}{N} \ln \Big ( 2^N \cosh( J \bar m + h )^N \Big)=
		  \ln  2  - \frac{J}{2} \bar m^2 + \ln \cosh( J \bar m + h )	
$$
which is the result.
		
\end{proof}

\subsection{Variational lower bound}

The second lower bound is provided by exploiting the well-known Gibbs entropic
variational principle (see \cite{ruelle}, pag. 188). In our case, instead of
considering the whole space of {\it ansatz} probability distributions considered in
\cite{ruelle}, we shall restrict to a much smaller one, and use the upper bound derived in the
last section in order to show that the lower bound corresponding to the restricted space
is sharp in the thermodynamic limit.

The mean-field nature of our Hamiltonian allows us to restrict the variational problem
to a product measure with only one degree of freedom, represented by the non-interacting
Hamiltonian:
$$
\tilde H=-r\sum_{i=1}^{N}\s_i,
$$
and so, given a Hamiltonian $\tilde H$, we define the ansatz Gibbs state corresponding to it as
$f(\sigma)$ as:
 \begin{equation*}
    \tilde\omega(f)=\frac{\sum_{\sigma}f(\sigma) e^{-\tilde H(\sigma)}}{\sum_{\sigma}e^{-\tilde H(\sigma)}}
    \end{equation*}

In order to facilitate our task, we shall express the variational principle of \cite{ruelle} in the
following simple form:
\begin{proposition}\label{varprin}
	Let a Hamiltonian $H$, and its associated partition function $\displaystyle{Z=\sum_{\sigma}e^{-H}}$
		be given. Consider an arbitrary trial Hamiltonian $\tilde H$ and its associated partition function $\tilde Z$.
		The following inequality holds:
			\begin{equation}\label{var ineq1}
				\ln Z \geqslant \ln \tilde Z - \tilde \omega (H) + \tilde {\omega} (\tilde H) \; .
			\end{equation}
		Given a Hamiltonian as defined in (\ref{Ham}) and its associated pressure per particle
		$p_N=\frac 1 N \ln Z$,
		the following inequality follows from {\rm (\ref{var ineq1})}:
			\begin{equation}\label{plow1}
				\liminf_{N\rightarrow \infty} p_N \geqslant \sup_{\bar m} p'_{low}
			\end{equation}
		\no where
			\begin{eqnarray}
				p'_{low}(\bar m)  &  = & \frac{J}{2} \bar m ^2 +  h \bar m
				- \frac{1+\bar m}{2}\ln (\frac{1+ \bar m}{2})  - \frac{1-\bar m}{2}\ln (\frac{1- \bar m}{2}).
			\end{eqnarray}
		\no and $\bar m \in [-1,1]$.
\end{proposition}

\begin{proof}
The inequality (\ref{var ineq1}) follows straightforwardly from Jensen's inequality:
	\begin{equation}
		e^{\tilde\omega(-H+\tilde H)} \le \tilde \omega (e^{-H+\tilde H}) \; .
	\end{equation}

We recall the Hamiltonian:
\begin{equation}\label{re_ham}
H(\sigma)=-\frac{J}{2N}\sum_{i,j}\s_i\s_j-h\sum_{i}\s_{i},
\end{equation}
\no so that its expectation on the trial state is
\begin{equation*}
    \tilde\omega(H)=-\frac{J}{2N}\sum_{i,j}\tilde\omega(\s_i\s_j)
    			       -h\sum_{i}\tilde\omega(\s_{i})
\end{equation*}

\no and a standard computation for the moments of a non-interacting system
(i.e. for a perfect gas) leads to
\begin{eqnarray}
    \tilde\omega(H) & = & -N (1-1/N) \frac{J}{2}(\tanh r) ^2- N\frac{J}{2} - N h\tanh r .
    \nonumber\\
\end{eqnarray}

Analogously, the trial Gibbs state of $\tilde H$ is:
\begin{equation*}
    \tilde\omega(\tilde H)  =   -N  r \, \tanh r,
\end{equation*}

\no and the non interacting partition function is:
$$
\tilde Z_N= \sum_{\sigma} e^{- \tilde H(\sigma)} =
2^{N}(\cosh r)^{N} ,
$$

\no which implies that the non-interacting pressure gives
$$
\tilde p_N = \frac 1 N \ln \tilde Z_N = \ln 2 +  \ln \cosh r
$$

So we can finally apply Proposition (\ref{var ineq1}) in order to find a lower bound for the pressure
$p_N=\displaystyle{\frac 1 N} \ln Z_N$:
\begin{eqnarray}
    p_N=\frac{1}{N}\ln Z_N  \geqslant \frac{1}{N}\   \Big( \ln \tilde Z_N
    - \tilde \omega (H) + \tilde {\omega} (\tilde H) \Big)
\end{eqnarray}

\no which explicitly reads:
\begin{eqnarray}\label{nonprecise}
    p_N=\frac{1}{N}\ln Z_N  & \geqslant &  \ln 2 + \ln \cosh r
    + \frac{J}{2} (\tanh r) ^2  +  h \tanh r    -  r  \tanh r
          \nonumber\\
    & &  + J / 2N - J  (\tanh r) ^2/N .
          \nonumber\\
\end{eqnarray}

Taking the liminf over $N$ and the supremum in $r$ of the left hand side we get (\ref{plow})
after performing the change of variables $\bar m=\tanh r$, and obtaining the following form
for the right hand side:

$$
p_{low}(\bar m)    = \frac{J}{2} \bar m ^2 +  h \bar m
				- \frac{1+\bar m}{2}\ln (\frac{1+ \bar m}{2})  - \frac{1-\bar m}{2}\ln (\frac{1- \bar m}{2}).
$$

\end{proof}

\subsection{Exact solution of the model}
	
We have derived two lower bounds and one upper bound to the thermodynamic
pressure, which are given by the suprema w.r.t. $\bar m$ of the following functions:
\begin{eqnarray}
 \nonumber
	p_{up}(\bar m)=p_{low}(\bar m) &=& \ln  2  - \frac{J}{2} \bar m^2 + \ln \cosh( J \bar m + h )
\\
	p'_{low}(\bar m)  &  = & \frac{J}{2} \bar m ^2 +  h \bar m
				- \frac{1+\bar m}{2}\ln (\frac{1+ \bar m}{2})  - \frac{1-\bar m}{2}\ln (\frac{1- \bar m}{2})
\label{plow2}
\end{eqnarray}

Since $p_{up}=p_{low}$, the supremum of this function gives the thermodynamic value of the
pressure, and thus provides the exact solution to the model.
However, it is important to verify that the bounds provided by all functions coincide,
since for more general cases one of the bounding arguments may fail, as indeed happens in the next chapter,
where a bound of type $p_{low}$ cannot be found due to lack of convexity in the Hamiltonian.
 Furthermore, $p'_{low}$
has a direct thermodynamic interpretation, as shall be explained in the following section.

For the standard Curie-Weiss model that we are studying here the equivalence of the two
bounds can be proved by way of a peculiar property of the Legendre transformation, and we
will do this in this section.

\begin{proposition}
The function
$$
f^*(y)= \frac{1}{J} \Big (  \frac{1+ y}{2}\ln \frac{1+y}{2} +   \frac{1-y}{2}\ln\frac{1-y}{2} - y\,h  \Big )
$$
is the Legendre transform of
$$
f(x)= \frac{1}{J} \ln \, 2 \, \cosh( Jx + h )
$$
\end{proposition}

\begin{proof}
The Legendre transformation is defined by
$$
f^*(y)=\sup_{x}\big (xy-f(x) \big)
$$

Since we are dealing with a convex function we can find the supremum by differentiation:
$$
\frac{df}{dx}=y-\tanh(Jx+h)=0
$$
which implies
$$
J x=\arctanh \,  y -h,
$$
so that by substituting we find that the Legendre transform of $f$ is
\begin{eqnarray*}
f^*(y)&=&y\, \frac{1}{J}(\arctanh \, y -h)-\frac{1}{J}\ln \, 2 \, \cosh( \, \arctanh \, y -h + h )=
\\
&=&y \, \frac{1}{J}\arctanh \, y - \frac{y h}{J} - \frac{1}{J} \ln \, 2 \, \cosh \, \arctanh \, y=
\\
&=& y \frac{1}{2J}\ln\frac{1+y}{1-y}-\frac{y h}{J} - \frac{1}{J} \ln \Big ( \exp \{ \frac{1}{2}\ln\frac{1+y}{1-y} \} +
 \exp \{ \frac{1}{2}\ln\frac{1-y}{1+y} \} \Big)=
\\
&=& y \frac{1}{2J}\ln\frac{1+y}{1-y}-\frac{y h}{J} - \frac{1}{J} \ln \Big (  \frac{1+y + 1 -y}{\sqrt{1-y^2}}  \Big)=
y \frac{1}{2J}\ln\frac{1+y}{1-y}- \frac{yh}{J} - \frac{1}{J}\ln \Big (  \frac{ 2 }{\sqrt{1-y^2}}  \Big)=
\\
&=&  \frac{1}{J} \Big ( \frac{1+ y}{2}\ln(1+y) +   \frac{1-y}{2}\ln(1-y) - y\,h - \ln 2 \Big ) =
\\
&=&   \frac{1}{J} \Big (  \frac{1+ y}{2}\ln \frac{1+y}{2} +   \frac{1-y}{2}\ln\frac{1-y}{2} - yh \Big ),
\end{eqnarray*}
which is the required result.

\end{proof}
We can similarly verify that the Legendre transform of $g(x)=-\frac{1}{2}x^2$ is
given by the function $g^*(x)=\frac{1}{2}x^2$.

This way we see that we can write the bounding functions as:
\begin{eqnarray}
 \nonumber
	p_{up}(\bar m)=p_{low}(\bar m) &=& J(f(\bar m)-g(\bar m)),
\\
	p'_{low}(\bar m)  &  = & J(g^*(\bar m)-f^*(\bar m)).
\end{eqnarray}
and the following proposition tells us that all of the bounds that we have found coincide.
\begin{proposition}
Let $f$ and $g$ be two convex functions and $f^*$ and $g^*$ be their Legendre transforms.
Then the following is true:
$$
\sup_x \, f(x) - g(x) = \sup_y \, g^*(y)-f^*(y)
$$
\end{proposition}

\begin{proof}

For a nice proof see \cite{ellis}, or the appendix in \cite{enter}.

\end{proof}
	
The last proposition tells us that both the variational principles we have derived provide
the correct value for the thermodynamic pressure, and so the results of this section
can be summarised in the following	
\begin{theorem}
Given a hamiltonian as defined in (\ref{Ham}), and defining the pressure per particle as
$\displaystyle{p_N=\frac 1 N \ln Z}$,
given parameters $J$ and $h$, the thermodynamic limit
$$
\lim_{N\rightarrow \infty} p_N = p
$$
\no of the pressure exists,
and can be expressed in one of the following equivalent forms:
      \newcounter{Lcount}
	\begin{list}{\alph{Lcount})}
	{\usecounter{Lcount}
	\setlength{\rightmargin}{\leftmargin}}
		\item $\displaystyle{p = \sup_{\bar m} \ p_{up}(\bar m)=\sup_{\bar m} \ p_{low}(\bar m)}$
		\item $\displaystyle{p = \sup_{\bar m} \ p'_{low}(\bar m)}$
    \end{list}
\end{theorem}

\section{Consistency equation}

In the last section we have expressed the thermodynamic pressure of the Curie-Weiss model
as the supremum of two distinct functions. Indeed, more can be said about this variational
principle, since even the argument of the supremum has a very important meaning: we shall see
in this section that, in case
there is a unique supremum for $p_{up}=p_{low}$ or $p'_{low}$, its argument gives the
thermodynamic value of the magnetization. If there exists more than one supremum, we have a phase
transition, and each argument gives a pure state for the magnetization.

First, we point out the straight-forward fact that stationary points of both
$p_{up}=p_{low}$ and $p'_{low}$ satisfy the condition:
\begin{equation}\label{mfeq}
	\bar m^*=\tanh(J\bar m^*+h),
\end{equation}
which can be found in the literature as {\it consistency equation}, {\it mean field equation},
{\it state equation}, {\it secularity equation}, and other names, depending on the context.

This equation is indeed important: since the bounding functions are smooth, and since it can
be easily seen by checking derivatives that none of the admit suprema at the boundary
of $[-1,1]$, we have as a consequence that any supremum of the function satisfies this equation.
It is also interesting to notice that the trivial fact that this equation has always  a solution
inside $[-1,1]$ can be also seen as a consequence of the existence results of Section
\ref{exists}.

\begin{proposition}\label{thmagn}
Let $J$ and $h$ be given so that $p_{up}=p_{low}$ has a unique supremum, which is
attained at $\bar m^*$. Then $\bar m^*=\lim_{N \ra \infty} \o_N(m)=\lim_{N \ra \infty} \o_N(\s_i)$.
\end{proposition}

\begin{proof}

The following holds at finite N, by definition of the pressure $p_N(J,h)$:
$$
\frac{\de p_N}{\de h}=\o_N(m_N).
$$

We have proved that $\{p_N\}$ is a convergent sequence of functions which are convex
(for a proof of the convexity of the pressure see \cite{golden}, where convexity is
proved for the free-energy in the Ising model, which is essentially the same as
the pressure multiplied by $-1$). This implies
that the limit function is also convex, and
as such it is differentiable almost everywhere. As a consequence we have the following:
$$
\lim_{N \ra \infty} \o_N(m)= \lim_{N \ra \infty} \frac{\de p_N}{\de h}=
\frac{\de \sup_{\bar m} p_{low}}{\de h}
$$
whenever the last derivative exists (for a proof that the limit of the derivatives
coincides with the derivative of the limit in this case see \cite{ellis} pag. 114).

Therefore if we write $\lim_{N \ra \infty}p_N=p(J,h,\bar m^*(J,h))$, we can write the following:
\begin{eqnarray*}
\frac{\de \sup_{\bar m} p_{low}}{\de h}= \frac{\de p(J,h,\bar m^*(J,h)}{\de h}
= -J\frac{\de \bar m^*}{\de h}\bar m^*+  \tanh(J \bar m^*+h)+J\frac{\de \bar m^*}{\de h}\tanh(J \bar m^*+h),
\end{eqnarray*}
and by substituting (\ref{mfeq}) we get
$$
\frac{\de \sup_{\bar m} p_{low}}{\de h}=\bar m^*,
$$
which is our result.

\end{proof}

A similar proposition can be proved analogously for $p'_{low}$. Let us now write
$$
\o(m)=\lim_{N \ra \infty} \o_N(m)      \quad  \textrm{and}  \quad   \o(\s_i)=\lim_{N \ra \infty} \o_N(\s_i).
$$
As a consequence of Proposition \ref{thmagn} we have that we can write
$$
p'_{low}(\bar m^*)=S-U
$$
where
$$
S=- \frac{1+\o(\s_i)}{2}\ln \Big (\frac{1+ \o(\s_i)}{2} \Big )  - \frac{1-\o(\s_i)}{2}\ln \Big ( \frac{1- \o(\s_i)}{2}\Big )
$$
is the thermodynamic entropy and
$$
U=\frac{J}{2} \o(m) ^2 +  h \o(m) 
$$
is the thermodynamic internal energy,
as can be derived directly from the Gibbs distribution.

\section{A heuristic approach}

We shall now describe a heuristic procedure to obtain the consistency equation
\ref{mfeq}.
First of all, we make the following observation about the Gibbs average $\o_N(\s_N)$
of the magnetization:
\begin{eqnarray*}
	\o_N(m)=\o_N(\s_1)=\frac{1}{Z_N}\sum_{\s \in \{-1,1\}^N } \s_1 e^{-H(\s)}
\end{eqnarray*}

We now define the following Hamiltonian $\tilde H_N$:
$$
\tilde H_N=-\frac{J}{2(N+1)}\sum_{i,j=1}^N\s_i\s_j - h\sum_{i=1}^N\s_i,
$$
and its associated partition function
$$
\tilde Z_N=\sum_{\s \in \{-1,1\} } e^{- \tilde H_N} ,
$$
which allows us to write:
\begin{eqnarray*}
	\o_N(\s_1)&=&\frac{\sum_{\s \in \{-1,1\}^N } \s_1  e^{\frac{J}{N}\sum_{i=1}^N \s_i \s_N + h \s_N }
										e^{- \tilde H_{N-1}(\s)} }
				{\sum_{\s \in \{-1,1\}^N }  e^{\frac{J}{N}\sum_{i=1}^N \s_i  \s_N + h \s_N }
										e^{- \tilde H_{N-1}(\s)}}=
\\
		&=&\frac{\tilde Z_N \sum_{\s \in \{-1,1\}^N } \s_1
					e^{\frac{J}{N}\sum_{i=1}^N \s_i \s_N  + h \s_N }
										e^{- \tilde H_{N-1}(\s)} }
				{\tilde Z_N \sum_{\s \in \{-1,1\}^N }  e^{\frac{J}{N}\sum_{i=1}^N \s_i \s_N + h \s_N }
										e^{- \tilde H_{N-1}(\s)}}=
\\
		&=&\frac{\tilde Z_N \sum_{\s \in \{-1,1\}^{N-1} } \s_1
					\sinh (\frac{J}{N}\sum_{i=1}^{N-1} \s_i + h + \frac{J}{N} )
										e^{- \tilde H_{N-1}(\s)}  }
				{\tilde Z_N \sum_{\s \in \{-1,1\}^{N-1} }
							 \cosh(\frac{J}{N}\sum_{i=1}^{N-1} \s_i  + h + \frac{J}{N} )
										e^{- \tilde H_{N-1}(\s) } }=
\\
		&=&\frac{\tilde \o_N(\sinh(\frac{J}{N}\sum_{i=1}^{N-1} \s_i + h + \frac{J}{N} ) ) }
				{\tilde \o_N(\cosh(\frac{J}{N}\sum_{i=1}^{N-1} \s_i + h + \frac{J}{N} ) )}
\end{eqnarray*}

Now, if we assume that the last line implies
\begin{equation}\label{assumcw}
\lim_{N \ra \infty} \o_N(\s_i)= \lim_{N \ra \infty} \frac{ \o_N(\sinh(J m+ h  ))}
							{ \o_N(\cosh(J m+ h  ) )}
\end{equation}
we can use the factorization properties of the model in order to derive the following.

Let us consider $\o_N(\sinh(J m+ h  ))$, and write it by making the power series at the argument
explicit:
$$
\o_N(\sinh(J m+ h  ))=\o_N  \Big (  \sum_{k=0}^{\infty} \frac{(J m+ h)^k}{(2k+1)!}  \Big )
$$

Now, if we consider only a partial sum up to $n$ at the argument of the Gibbs state, and take the thermodynamic
limit, the self-averaging property of the magnetization tells us that the following holds a.e. in $J$ and $h$:

\begin{eqnarray*}
	\lim_{N \ra \infty} \o_N  \Big (  \sum_{k=0}^{n} \frac{(J m+ h)^k}{(2k+1)!}  \Big )
       &=&\lim_{N \ra \infty} \o_N  \Big (  \sum_{k=0}^{n} \frac{1}{(2k+1)!} \sum_{l=0}^{k } {k \choose l}  (J m)^l h^{k-l}  \Big )=
 \nonumber \\
       &=&\lim_{N \ra \infty}   \sum_{k=0}^{n} \frac{1}{(2k+1)!} \sum_{l=0}^{k } {k \choose l}  J^l \o_N (m)^l h^{k-l}  \Big )=
\nonumber \\
        &=&\lim_{N \ra \infty}   \sum_{k=0}^{n} \frac{(J  \o_N(m)+ h)^k}{(2k+1)!}
\label{summom}
\end{eqnarray*}

Now, disregarding convergence problems, the limit of (\ref{summom}) together with the
assumption  (\ref{assumcw}) give the following equation:
$$
\bar m^*=\tanh(J \bar m^*+h),
$$
where $\bar m^*$ is the thermodynamic magnetization. This way we have derived heuristically the consistency equation describing the most important quantity for our
model just by making use of the model's factorization properties.

It is important, however, to stress that the procedure we proposed in this section is not mathematically rigorous: assumption (\ref{assumcw}), though sensible, hasn't been derived
rigorously, and the possible convergent problems have not been considered.
Nevertheless, since the procedure has provided the right answer which we have
derived rigorously throughout the chapter, and since it consists simple considerations,
it can be see as a way of approaching models
defined on random networks instead that on the complete graph,
which are not as well understood as the one
treated in this chapter.
%
%

%% file: capitolo_cw2d.tex
\resettheoremcounters

\chapter{The Curie-Weiss model for many populations}\label{manycw}

In this chapter we consider the problem of characterizing the equilibrium statistical mechanics
of an mean field interacting system partitioned into $p$ sets of spins. The relevance of such a problem
to social modelling
is that such a partition can be made to correspond to the partition into classes of people sharing
the same socio-economics attributes, as described in chapter \ref{ds}.


Our results can be summarised as follows. After introducing the model we show in section
3 that it is well posed by showing that its thermodynamic limit exists. The result is non-trivial,
since sub-additivity is not met at finite volume. In section 4 we show that the system
fulfills a factorization property for the correlation functions which reduces the equilibrium state
to only $p$ degrees of freedom. The method is conceptually similar to the one developed by
Guerra in \cite{aod} to derive identities for the overlap distributions in the Sherrington and Kirkpatrick
model.

We also derive the pressure of the model by rigorous methods developed in the recent
study of mean field spin glasses (see \cite{guerrarev} for a review). It is interesting to
notice that though very simple,
our model encompasses a range of regimes that do not admit solution by the
elegant interpolation method used in the celebrated existence result of the Sherrington and Kirkpatrick model \cite{guerra}.
This is due to the lack of positivity of the quadratic form describing
the considered interaction.
Nevertheless we are able to solve the model exactly in section 4.4, using the lower bound provided by the Gibbs
variational principle, and thanks to a further bound given by
a partitioning of the configuration space, itself originally devised in the study of spin glasses
(see \cite{guerrarev, desanctis, guerrarev2}).

As in the classical Curie-Weiss model, the exact solution is provided in an implicit form;
for our system, however, we find a system of equations of state, which are coupled as well as trascendental,
and this makes the full characterization of all the possible regimes highly non-trivial.
A simple analytic result about the number of solutions for the two-population case is
proved in section \ref{maxima}.


\section{The Model}\label{model1}

We can generalize the Curie-Weiss model to $p$-populations, allowing
$r$-body interactions with $r=1..p$. This gives rise to the following Hamiltonian:
\begin{equation}\label{generalham}
H_N=-N\sum_{r=1}^{p} \sum_{i_1,..., i_r=1}^{p} J_{i_1,...,i_r} \prod_{k=1}^{r} m_{i_k},
\end{equation}
or, equivalently, to the following Utility function for individual $i$:
$$
U_i=\sum_{r=1}^{p} \sum_{i_1,..., i_{r-1}=1}^{p} J_{i_1,...,i_{r-1}, i} \prod_{k=1}^{r-1} m_{i_k}.
$$

Here $J_{i_1,...,i_r}$ gives the interaction coefficients corresponding to the $r$-body interaction
among individuals coming from populations $i_1,...,i_r$, respectively.
We can also consider the external fields to be already included in this form of the model,
just by setting $J_i=h_i$. So we have defined
interactions by using a tensor $J_{i_1,...,i_r}$ of rank  $r$ for each of the $r$-body interactions.

\section{Existence of the thermodynamic limit for many populations}

We shall prove that our model admits a thermodynamic limit by exploiting an existence theorem
provided for mean field models in \cite{bcg}: the result states that the existence of the pressure
per particle for large volumes is guaranteed by a monotonicity condition on the equilibrium
state of the Hamiltonian. Such a result proves to be quite useful when the condition of
convexity introduced by the interpolation method \cite{guerra,guerrarev} doesn't apply due to
lack of positivity of the quadratic form representing the interactions.
%
We therefore prove the existence of the thermodynamic limit independently of an exact solution. Such
a line of enquiry is pursued in view of further refinements of our model, that shall possibly involve
random interactions of spin glass or random graph type, and that might or might not come with
an exact expression for the pressure.
\begin{proposition}\label{existence}
There exists a function $p$ of all the parameters $J_{i_1,...,i_r}$ such that
$$
\lim_{N \rightarrow \infty} p_N= p \; .
$$
\end{proposition}
\vskip .1 in
The previous proposition is proved with a series of lemmas.
Theorem 1 in \cite{bcg} states that given a Hamiltonian $H_N$ and its associated equilibrium state
$\omega_N$ the model admits a thermodynamic limit whenever the physical condition
\begin{equation}\label{additivity}
\omega_N (H_N) \geqslant \omega_N (H_{N_1}) + \omega_N (H_{N_2}), \qquad N_1 + N_2= N,
\end{equation}
\no is verified.

We proceed by first verifying this condition for an alternative Hamiltonian $\tilde H_N$, and then showing
that its pressure $\tilde p_N$ tends to our original pressure $p_N$ as $N$ increases. We choose $\tilde H_N$
in such a way that the condition (\ref{additivity}) is verified as an equality.

Now, define the alternative Hamiltonian $\tilde H_N$ as follows:
$$
\tilde H_N = - C \, N \,  \prod_{l=1}^p \frac{(N_{i_l}-k_l)!}{N_{i_l}!}
             \sum_{j_k=N_{i_k-1}+1,...,N_{i_k} \atop  j_k \neq j_h \textrm{ for } k \neq h}  \s_{j_1}...\s_{j_r}
$$
where $C$ is a real number.

Though the notation is cumbersome at this point, the new Hamiltonian simply considers products of $r$ distinct
spins, $k_i$ of which are taken from population $i$ (i.e. $\sum_{i=1}^{p}k_i=r$) and so the combinatorial coefficient
is just dividing the sum by the correct number of terms contained in the sum itself.


\begin{lemma}\label{workinglimit}
There exists a function $\tilde p$ such that
$$
\lim_{N \rightarrow \infty} \tilde p_N= \tilde p
$$
\end{lemma}

\vskip .1 in
\begin{proof}

By linearity we have that
\begin{equation}\label{adty}
\o_N(\tilde H_N)=- C \, N \,  \prod_{l=1}^p \frac{(N_{i_l}-k_l)!}{N_{i_l}!}
             \sum_{j_k=N_{i_k-1}+1,...,N_{i_k} \atop  j_k \neq j_h \textrm{ for } k \neq h} \o_N( \s_{j_1}...\s_{j_r} ) =
             -C \, N \, \o_N(\s_{j_1}...\s_{j_r}),
\end{equation}\label{adty}
where, with a little abuse of notation, we let $\s_{j_1},..,\s_{j_r}$, after the last equality be distinct spins taken
from their own respective populations. The last equality hence follows from the invariance of $\tilde H_N$ with respect to
permutations of spins belonging to the same population.

Equation (\ref{adty}) implies trivially
$$
\o_N(\tilde H_N - \tilde H_{N_1} - \tilde H_{N_2})=0
$$
for $N_1+N_2=N$, which verifies (\ref{additivity}) as an equality.

\end{proof}

The following two Lemmas show that the difference between $H_N$ and $\tilde H_N$ is
thermodynamically negligible and as a consequence their pressures coincide in the thermodynamic
limit.

Though the notation is quite tedious, the proof is in no way different from the one
described in \cite{bcg}. We chose to keep full generality during this existence proof in order to
show that the mean-field allows one to consider a whole range of possibilities for interaction,
which might turn out useful for the modelling effort.

\begin{lemma}\label{order}

\begin{equation}
H_N = \tilde H_N + O(1)
\end{equation}

\no i.e.
$$
\lim_{N \rightarrow \infty}  \frac{H_N}{N} = \lim_{N \rightarrow \infty} \frac{\tilde H_N}{N}
$$

\end{lemma}

\begin{proof}

We begin the proof by rephrasing the Hamiltonian in term of the spins, as follows:
\begin{eqnarray*}
    H_N &=& -N\sum_{r=1}^{p} \sum_{i_1,..., i_r=1}^{p} \Big\{ J_{i_1,...,i_r} \prod_{k=1}^{r} \frac{ N_{i_k} }{ N_{i_k} } m_{i_k} \Big\}=
\\
        &=& - \sum_{r=1}^{p} \Big\{ \sum_{i_1,..., i_r=1}^{p} N \frac{ N^r }{ N^r } \prod_{k=1}^{r} \frac{ 1 }{ N_{i_k} } J_{i_1,...,i_r} \prod_{k=1}^{r}
                            N_{i_k} m_{i_k} \Big\} =
\\
        &=& - \sum_{r=1}^{p} \sum_{i_1,..., i_r=1}^{p} \Big\{ \frac{1}{ N^{r-1} } \prod_{k=1}^{r} \frac{ 1 }{ \a_{i_k} }
            J_{i_1,...,i_r} \sum_{j_k=N_{i_k-1}+1,...,N_{i_k}} \s_{j_1}...\s_{j_r} \Big\} =
\end{eqnarray*}
where
$$
N=\sum_{i=1}^p N_i, \quad \a_i=\frac{N_i}{N}, \quad N_0=0.
$$

We only need to give details of the proof in the case only one of the coefficients $J_{i_1,...,i_r}\neq0$.
The general case follows by summing up all the terms corresponding to non-zero interacting coefficients and noticing
that, since this sum has only finitely many terms, the result still holds.

So we consider the following Hamiltonian
$$
H_N=-NJ_{i_1,...,i_r}\prod_{k=1}^{r} m_{i_k}=\frac{1}{ N^{r-1} } \prod_{k=1}^{r} \frac{ 1 }{ \a_{i_k} }
            J_{i_1,...,i_r}  \sum_{j_k=N_{i_k-1}+1,...,N_{i_k}} \s_{j_1}...\s_{j_r} ,
$$
and we can lighten our notation by setting $C=\frac{ 1 }{ \a_{i_k} }J_{i_1,...,i_r}$,
$$
H_N=\frac{C}{ N^{r-1} } \sum_{j_k=N_{i_k-1}+1,...,N_{i_k}} \s_{j_1}...\s_{j_r}.
$$

Now, following \cite{bcg} we divide the sum in two parts, as follows:
$$
H_N=\frac{C}{ N^{r-1} } \sum_{j_k=N_{i_k-1}+1,...,N_{i_k} \atop  j_k \neq j_l \textrm{ for } k \neq l}  \s_{j_1}...\s_{j_r}+
    \frac{C}{ N^{r-1} } \sum^* \s_{j_1}...\s_{j_r}.
$$

The first part is a sums only over products of distinct spins, whereas $\sum^*$ is a sum of all products where
at least two spins are equal. It is straightforward to show that
$$
    \frac{C}{ N^{r-1} } \sum^* \s_{j_1}...\s_{j_r}=O(1),
$$
so that we can rewrite $H_N$ as follows:
$$
H_N=\frac{C}{ N^{r-1} } \sum_{j_k=N_{i_k-1}+1,...,N_{i_k} \atop  j_k \neq j_l \textrm{ for } k \neq l}  \s_{j_1}...\s_{j_r}+
    O(1).
$$

A straightforward calculation comparing $H_N$ and $\tilde H_N$ can now check that
$$
H_N=\tilde H_N + O(1),
$$
which is our result.

\end{proof}

\begin{lemma}\label{presham}
Say $\ds{p_N=\frac{1}{N} \ln Z_N }$, and say $h_N(\sigma)=\ds{\frac{H_N(\sigma)}{N}   }$.
Define $\tilde Z$, $\tilde p_N$ and $\tilde h_N$ in an analogous way.

Define
\begin{equation}\label{norm}
	k_N = \ds{\|h_N-\tilde h_N \|=\sup_{\sigma \in \{-1,+1\}^{N}} \{ |h_N(\sigma)- \tilde h_N(\sigma) | \} < \infty  }.
\end{equation}

Then
$$
|p_N- \tilde p_N| \leqslant \|h_N- \tilde h_N\| \; .
$$
\end{lemma}

\begin{proof}
\begin{eqnarray*}
p_N-\tilde p_N&=& \frac{1}{N} \ln Z_N - \frac{1}{N} \ln \tilde Z_N = \frac{1}{N} \ln \frac{Z_N}{\tilde Z_N}
\\
&=& \frac{1}{N} \ln \frac{\sum_{\sigma} e^{-H_N(\sigma)} }{\sum_{\sigma}e^{-\tilde H_N(\sigma)}}
\leqslant \frac{1}{N} \ln \frac{\sum_{\sigma} e^{-H_N(\sigma)}}{\sum_{\sigma}e^{-N(h_N(\sigma) + k_N)}}=
\\
&=& \frac{1}{N} \ln \frac{\sum_{\sigma} e^{-H_N(\sigma)}}{e^{-Nk_N}\sum_{\sigma}e^{-N h_N(\sigma)}}= \frac{1}{N} \ln e^{Nk_N}
= k_N = \|h_N-\tilde h_N\|
\end{eqnarray*}

\no where the inequality follows from the definition of $k_N$ in (\ref{norm}) and from monotonicity of the exponential and logarithmic functions.
The inequality for $\tilde p_N-p_N$ is obtained in a similar fashion.
\end{proof}

We are now ready to prove the main result for this section:

\begin{proofof}{Proposition \ref{existence}}
The existence of the thermodynamic limit follows from our Lemmas. Indeed, since
by Lemma \ref{workinglimit} the limit for $\tilde p_N$ exists, Lemma \ref{presham}
and Lemma \ref{order} tell us that
$$
\lim_{N \rightarrow \infty} |p_N-\tilde p_N | \leqslant \lim_{N \rightarrow \infty} \|h_N-\tilde h _N \| = 0,
$$
\no implying our result.
\end{proofof}

\section{Factorization properties}

From now on we shall restrict the model to include pair interactions only. Therefore, we have a Hamiltonian
of the following kind:
\begin{equation}\label{ham1}
    H_N=-N \sum_{i,j=1}^{p} \frac{J_{i,j}}{2} m_{i}  m_{j} - N \sum_{i=1}^{p} h_{i} m_{i} ,
\end{equation}

In this section we shall prove that the correlation functions of our model factorize completely
in the thermodynamic limit, for almost every choice of parameters. This implies that
all the thermodynamic properties of the system can be described by the magnetizations
$m_i$ of the $p$ populations defined in Section \ref{model1}. Indeed, the exact solution
of the model,
to be derived in the next section, comes as $p$ coupled equations of state for the $m_i$.

\begin{proposition}
$$
\lim_{N \rightarrow \infty} \big ( \omega_N(\sigma_i\sigma_j)-\omega_N(\sigma_i)\omega_N(\sigma_j) \big ) =0
$$

\no for almost every choice of parameters, where $\sigma_i$, $\sigma_j$
are any two distinct spins in the system.
\end{proposition}

\begin{proof}
We recall the definition of the Hamiltonian
$$
H_N=-N \sum_{i,j=1}^{p} J_{i,j} m_{i}  m_{j} - N \sum_{i=1}^{p} h_{i} m_{i} ,
$$
and of the pressure per particle
$$
p_N=\frac{1}{N} \ln \sum_{\sigma} e^{-H_N(\sigma)}.
$$

By taking first and second partial derivatives of $p_N$ with respect to $h_i$ we get
$$
\frac{\partial p_N}{\partial h_i}=\frac{1}{N} \sum_{\sigma} N m_i(\sigma) \frac{e^{-H(\sigma)}}{Z_N}= \omega_N(m_i),
\qquad
\frac{\partial^2 p_N}{\partial \; h_i^2}=N  ( \omega_N(m_i^2)-\omega_N(m_i)^2).
$$

By using these relations we can bound above the integral with respect to $h_i$ of the fluctuations of $m_i$ in the Gibbs state:
\begin{eqnarray}\label{average var}
\Bigg |\int_{h_i^{(1)}}^{h_i^{(2)}} ( \omega_N(m_i^2)-\omega_N(m_i)^2 ) \; dh_i \Bigg | & = &
\frac{1}{N } \Bigg | \int_{h_i^{(1)}}^{h_i^{(2)}}   \frac{\partial^2 p_N}{\partial h_i^2}   \; dh_i \Bigg | =
\frac{1}{N } \Bigg | \int_{h_i^{(1)}}^{h_i^{(2)}} \frac{\partial p_N}{\partial \; h_i} \bigg |_{h_i^{(1)}}^{h_i^{(2)}} \Bigg | \leqslant
\nonumber  \\
& \leqslant & \frac{1}{N } \big ( \big | \omega_N(m_i)|_{h_i^{(2)}} \big | + \big | \omega_N(m_i)|_{h_i^{(1)}} \big | \big ) = O \big (\frac{1}{N} \big ).
\nonumber  \\
\end{eqnarray}

On the other hand we have that
$$
\omega_N(m_i)=\frac{\partial p_N}{\partial h_i},
$$
and
$$
\omega_N(m_i^2)=2\frac{\partial p_N}{\partial J_{i,i}},
$$
so, by convexity of the thermodynamic pressure $\ds p=\lim_{N \rightarrow \infty} p_N$,
both quantities $\ds \frac{\partial p_N}{\partial h_i}$ and $\ds \frac{\partial p_N}{\partial J_{i,i}}$ have well
defined thermodynamic limits almost everywhere.
This together with (\ref{average var}) implies that
\begin{equation}\label{ae m1}
\lim_{N \rightarrow \infty} ( \omega_N(m_i^2)-\omega_N(m_i)^2 ) = 0 \quad \textrm{a.e. in $h_i$, $J_{i,i}$}.
\end{equation}

In order to prove our statement we shall write the magnetization $m_i$ in terms of spins belonging to
the $i^{th}$ population, and then use the permutation invariance of the Gibbs measure:
\begin{eqnarray}\label{xi i xi j}
\omega_N(m_i)&=&\omega_N(\frac{1}{N_i-N_{i-1}}\sum_{j=N_{i-1}}^{N_i} \s_i )= \omega_N(\s_1),
\nonumber \\
\omega_N(m_i^2)&=&\omega_N(\frac{1}{(N_i-N_{i-1})^2}\sum_{j,\ l=N_{i-1}}^{N_i} \s_j \s_l )=
\nonumber \\
&=& \omega_N(\frac{1}{(N_i-N_{i-1})^2}\sum_{j\neq l=N_{i-1}}^{N_i} \s_j \s_l )+
\omega_N(\frac{1}{(N_i-N_{i-1})^2}\sum_{j = l=N_{j-1}}^{N_j} \s_j \s_l ) =
\nonumber \\
&=& \frac{N_i-N_{i-1}-1}{N_i-N_{i-1}} \ \omega_N(\s_1 \s_2) + \frac{1}{N_i-N_{i-1}}. \nonumber
\\
\end{eqnarray}

We have that (\ref{xi i xi j}) and (\ref{ae m1}) imply
\begin{equation}\label{part one}
\lim_{N \rightarrow \infty} \omega_N(\s_i\s_j)-\omega_N(\s_i)\omega_N(\s_j)=0,
\end{equation}
\no which verifies our statement for all couples of spins $i \neq j$ belonging to
the same population.

Furthermore, by defining $\Var_N(m_i)=\big (\omega_N(m_i^2)-\omega_N(m_i)^2 \big )$ for all populations $i$, we exploit
(\ref{ae m1}), and
use the Cauchy-Schwartz inequality to get
\begin{equation}\label{ae m1 m2}
|\omega_N(m_i m_j)-\omega_N(m_i)\omega_N(m_j)| \leqslant \sqrt{\Var(m_i)\Var(m_j)} \ {\ds{ \ulimit{\footnotesize N \rightarrow \infty}}} \ 0
\quad \textrm{a.e. in $J_{i,i}$, $J_{j,j}$, $h_i$, $h_j$}
\end{equation}

By using (\ref{xi i xi j}) and (\ref{ae m1 m2}) we can therefore verify
statements which are analogous to (\ref{part one}), but which concern
$\omega_N(\s_i \s_j)$ where $\s_i$ and $\s_j$ are spins belonging to
different subsets.

We have thus proved our claim for any couple of spins in the global system.

\end{proof}

\section{Solution of the model}


We shall derive upper and lower bounds for the thermodynamic limit of the pressure. The lower bound is obtained through the standard entropic variational principle, while the upper bound
is derived by a decoupling strategy.
%

\subsection{Upper bound}

In order to find an upper bound for the pressure we shall divide the configuration space into a partition of
microstates of equal magnetization, following \cite{desanctis, guerrarev, guerrarev2}. Since
each population $g$ consists of $N_g$ spins, its magnetization can take
exactly $N_g+1$ values, which are the elements of the set
$$
R_{N_g}=\Big \{-1, -1+\frac {1} {2 N_g}, \dots ,1 -\frac {1} {2 N_g}, 1 \Big \}.
$$

Clearly for every $m_g(\sigma)$ we have that
$$
\sum_{\bar m_g \in R_{N_g}} \delta_{m_g, \bar m_g}=1,
$$
\no where $\delta_{x,y}$ is a Kronecker delta. This allows us to rewrite the partition function as follows:
\begin{eqnarray} \label{deltas}	
Z_N &= &\sum_\sigma   \exp \big\{ \frac{N}{2} \sum_{i,j=1}^{p} J_{i,j} m_{i}  m_{j} + N \sum_{i=1}^{p} h_{i} m_{i} \big \} ={}
\nonumber \\
     & = & \sum_\sigma \sum_{\forall g \; \bar m_g \in R_{N_g}}
          \prod_{g=1}^p \delta_{m_g, \bar m_g}
   \exp\big\{\frac{N}{2} \sum_{i,j=1}^{p} J_{i,j} m_{i}  m_{j} + N \sum_{i=1}^{p} h_{i} m_{i} \big \}.
\end{eqnarray}

Thanks to the Kronecker delta symbols,
we can substitute $m_i$ (the average of the spins within a configuration)
with the parameter $\bar m_i$ (which is not coupled to the spin configurations)
in any convenient fashion.

Therefore we can use the following relations in order to linearize all quadratic terms
appearing in the Hamiltonian
\begin{eqnarray*}
(m_i - \bar m_i)^2 & = & 0 \ \forall i,
\nonumber\\
(m_i - \bar m_i)(m_j - \bar m_j) & = & 0 \ \forall i \neq j,.
\end{eqnarray*}


Once we've carried out these substitutions into (\ref{deltas}) we are left with a function
which depends only linearly on the $m_i$:
\begin{eqnarray*}
Z_N & = & \sum_\sigma \sum_{\forall g \; \bar m_g \in R_{N_g}}
          \prod_{g=1}^p \delta_{m_g, \bar m_g}
   \exp\big\{ \frac{N}{2} \sum_{i,j=1}^{p} J_{i,j} m_{i}  m_{j} + N \sum_{i=1}^{p} h_{i} m_{i} \big \} =
\\
 & = & \sum_\sigma \sum_{\forall g \; \bar m_g \in R_{N_g}}
          \prod_{g=1}^p \delta_{m_g, \bar m_g}
   \exp\big\{ \frac{N}{2} \sum_{i,j=1}^{p} J_{i,j} (m_{i}  \bar m_{j} + \bar m_{i} m_{j} - \bar m_{i}  \bar m_{j})
                                                               + N \sum_{i=1}^{p} h_{i} m_{i} \big \} =
\\
 & = & \sum_\sigma \sum_{\forall g \; \bar m_g \in R_{N_g}}
          \prod_{g=1}^p \delta_{m_g, \bar m_g}
   \exp\big\{ - \frac{N}{2} \sum_{i,j=1}^{p} J_{i,j} \bar m_{i}  \bar m_{j} +
                            \frac{N}{2} \sum_{i,j=1}^{p} J_{i,j} (m_{i}  \bar m_{j} + \bar m_{i} m_{j}) +
\\
&&    + N \sum_{i=1}^{p} h_{i} m_{i} \big \} =
\end{eqnarray*}
\no and bounding above the Kronecker deltas by $1$ we get

\begin{eqnarray} \label{deltas2}
Z_N & \leqslant & \sum_\sigma \sum_{\forall g \; \bar m_g \in R_{N_g}}
   \exp\big\{ \frac{N}{2} \sum_{i,j=1}^{p} J_{i,j} \bar m_{i}  \bar m_{j} +
                            \frac{N}{2} \sum_{i,j=1}^{p} J_{i,j} (m_{i}  \bar m_{j} + \bar m_{i} m_{j}) +
                                                                N \sum_{i=1}^{p} h_{i} m_{i} \big \} =
\nonumber\\
\end{eqnarray}

As observed many times by Guerra \cite{guerrarev}, since both sums are taken over finitely many terms,
it is possible to exchange the order of the
two summation symbols, in order to carry out the sum over the spin configurations, which now
factorizes, thanks to the linearity of the interaction with respect to the $m_g$. This way we get:
$$
Z_N \leqslant \sum_{\forall g \; \bar m_g \in R_{N_g} } G(\bar m_1, ... , \bar m_p).
$$
where
\begin{eqnarray}\label{g}
G &=&  \exp\big\{- \frac{N}{2} \sum_{i,j=1}^{p} J_{i,j} \bar m_{i}  \bar m_{j}\}
   \cdot \prod_{j=1}^p 2^{N_j}\big( \cosh \big( \sum_{i=1}^p \frac{J_{i,j}+J_{j,i}}{2 \a_j} \bar m_i+  \frac{h_j}{\a_j} \big) \big) ^{N_j}\nonumber\\
\end{eqnarray}
where
$$
\a_j=\frac{N_j}{N}
$$

Since the summation is taken over the
ranges   $R_{N_g}$, of cardinality $N_g+1$, we get that the total number of terms
is $\ds \prod_{g=1}^{p}(N_g+1)$. Therefore
\begin{eqnarray}
Z_N & \leqslant   & \prod_{g=1}^p (N_g+1) \sup_{\bar m_1, ... , \bar m_p } G,
 \end{eqnarray}

\no which leads to the following upper bound for $p_N$:
\begin{eqnarray} \label{deltas3}
p_N & = & \frac{1}{N} \ln Z_N  \leqslant \sum_{g=1}^p \frac{1}{N} \ln (N_g+1) + \frac{1}{N} \ln \sup_{\bar m_1,..., \bar m_p} G \; .
\end{eqnarray}

Now defining the $N$ independent function
\begin{eqnarray}
p_{UP}=\frac{1}{N} \ln G  & =  & \ln 2 - \frac{1}{2} \sum_{i,j=1}^{p} J_{i,j} \bar m_{i}  \bar m_{j} +
\sum_{j=1}^p \a_j\ln \cosh \big( \sum_{i=1}^p \frac{J_{i,j}+J_{j,i}}{2 \a_j} \, \bar m_i+  \frac{h_j}{\a_j} \big),
\nonumber\\
\end{eqnarray}
where
$$
\a_j=\frac{N_j}{N}
$$

\no the thermodynamic limit gives:
\begin{equation}
\limsup_{N \rightarrow \infty} p_N \leqslant  \sup_{\bar m_1, ... , \ \bar m_p} p_{UP}.
\end{equation}

We can summarize the previous computation into the following:

\begin{lemma}
Given a Hamiltonian as defined in (\ref{ham1}), and defining the pressure per particle as $p_N=\frac 1 N \ln Z$,
given parameters $J_{i,j}$ and $h_i$,
the following inequality holds:
$$
				\limsup_{N\rightarrow \infty} p_N \leqslant \sup_{\bar m_1, ... , \ \bar m_p} p_{UP}
$$
\no where
\begin{eqnarray}
p_{UP} & =  & \ln 2 -\frac{1}{2} \sum_{i,j=1}^{p} J_{i,j} \bar m_{i}  \bar m_{j} +
            \sum_{j=1}^p \a_j\ln \cosh \big( \sum_{i=1}^p \frac{J_{i,j}+J_{j,i}}{2 \, \a_j} \, \bar m_i+  \frac{h_j}{\a_j} \big),
\end{eqnarray}
\no and $\bar m_i \in [-1,1]$.
\end{lemma}

\subsection{Lower bound}

The lower bound is provided by exploiting the well-known Gibbs entropic
variational principle (see \cite{ruelle}, pag. 188). In our case, instead of
considering the whole space of {\it ansatz} probability distributions considered in
\cite{ruelle}, we shall restrict to a much smaller one, and use the upper bound derived in the
last section in order to show that the lower bound corresponding to the restricted space
is sharp in the thermodynamic limit.

The mean-field nature of our Hamiltonian allows us to restrict the variational problem
to a p-degrees of freedom product measures represented through the non-interacting
Hamiltonian:
$$
\tilde H=-r_1\sum_{i=1}^{N_1}\s_i-r_2\sum_{i=N_1+1}^{N_1+N_2}\s_i+...-r_p\sum_{i=\sum_{i=1}^{p-1}N_i+1}^{N}\s_i,
$$
%
and so, given a Hamiltonian $\tilde H$, we define the ansatz Gibbs state corresponding to it as
$f(\sigma)$ as:
 \begin{equation*}
    \tilde\omega(f)=\frac{\sum_{\sigma}f(\sigma) e^{-\tilde H(\sigma)}}{\sum_{\sigma}e^{-\tilde H(\sigma)}}
    \end{equation*}

In order to facilitate our task, we shall express the variational principle of \cite{ruelle} in the
following simple form:

\begin{proposition}\label{varprin}
Let a Hamiltonian $H$, and its associated partition function $\displaystyle{Z=\sum_{\sigma}e^{-H}}$
be given. Consider an arbitrary trial Hamiltonian $\tilde H$ and its associated partition function $\tilde Z$.
The following inequality holds:
\begin{equation}\label{var ineq}
\ln Z \geqslant \ln \tilde Z - \tilde \omega (H) + \tilde {\omega} (\tilde H) \; .
\end{equation}
Given a Hamiltonian as defined in (\ref{ham}) and its associated pressure per particle
$p_N=\frac 1 N \ln Z$,
the following inequality follows from {\rm (\ref{var ineq})}:
\begin{equation}\label{plow}
\liminf_{N\rightarrow \infty} p_N \geqslant \sup_{\bar m_1, ... ,\bar m_p} p_{LOW}
\end{equation}
\no where
\begin{eqnarray}
p_{LOW}  &  = & \frac{1}{2}  \sum_{g, k =1}^p   J_{g,k} \bar m_g \bar m_k  + \sum_{g=1}^p h_g \bar m_g
            +\sum_{g=1}^p \a_g S(\bar m_g),
\end{eqnarray}
the function $S( \bar m_g)$ being the entropy
$$
S(\bar m_g)= - \frac{1+\bar m_g}{2}\ln (\frac{1+ \bar m_g}{2}) - \frac{1-\bar m_g}{2}\ln (\frac{1- \bar m_g}{2})
$$
\no and $\bar m_g \in [-1,1]$.
\end{proposition}

\begin{proof}
The (\ref{var ineq}) follows straightforwardly from Jensen's inequality:
\begin{equation}
e^{\tilde\omega(-H+\tilde H)} \le \tilde \omega (e^{-H+\tilde H}) \; .
\end{equation}

The Hamiltonian (\ref{ham1}) can be written in term of spins as:
\begin{equation}\label{spin_ham}
H(\sigma)=-\frac{1}{2N}  \sum_{g,k=1}^p  \Big \{ \frac{J_{g,k}}{\a_g \a_k}
\sum_{i \in P_g , \ j \in P_k}\s_i\s_j \Big \}-
\sum_{g=1}^p \{ \frac{h_g}{\a_g}\sum_{i \in P_g}\s_{i}\}, \; ;
\end{equation}
where $P_g$ contains the labels for spins belonging to the $g^{th}$ subpopulation, that is
$$
P_g=\{\sum_{k=1}^{g-1}N_{k}+1,\sum_{k=1}^{g-1}N_{k}+2,...,\sum_{k=1}^{g}N_{k}\}
$$

\no indeed its expectation on the trial state is
\begin{equation}
    \tilde\omega(H)=-\frac{1}{2N}  \sum_{g,k=1}^p  \Big \{ \frac{J_{g,k}}{\a_g \a_k}
\sum_{i \in P_g , \ j \in P_k} \tilde\omega(\s_i\s_j) \Big \}-
\sum_{g=1}^p \{ \frac{h_g}{\a_g}\sum_{i \in P_g}\tilde\omega(\s_{i})\}
\end{equation}

\no and a standard computation for the moments leads to

\begin{eqnarray}
    \tilde\omega(H) & = & -\frac{N}{2}  \sum_{g=1}^p  (1-\frac{1}{N\a_g}) J_{g,g} (\tanh r_g) ^2
   -\frac{1}{2}  \sum_{g=1}^p \frac{1}{\a_g J_{g,g}}
    -\frac{N}{2}  \sum_{g \neq k =1}^p   J_{g,k} \tanh r_g \tanh r_k
    \nonumber\\
 & &  - N \sum_{g=1}^p h_g \tanh r_g.
    \nonumber\\
\end{eqnarray}

Analogously, the Gibbs state of $\tilde H$ is:
\begin{equation*}
    \tilde\omega(\tilde H)  =   -N \sum_{g=1}^p \alpha_g \, r_g \, \tanh r_g,
\end{equation*}

\no and the non interacting partition function is:
$$
\tilde Z_N= \sum_{\sigma} e^{- \tilde H(\sigma)} =
\sum_{g=1}^p 2^{N_g}(\cosh \, r_g)^{N_g}
$$

\no which implies that the non-interacting pressure gives
$$
\tilde p_N = \frac 1 N \ln \tilde Z_N = \ln 2 + \sum_{g=1}^p \a_g \ln \cosh \, r_g
$$

So we can finally apply Proposition (\ref{var ineq}) in order to find a lower bound for the pressure
$p_N=\displaystyle{\frac 1 N} \ln Z_N$:
\begin{eqnarray}
    p_N=\frac{1}{N}\ln Z_N  \geqslant \frac{1}{N}\   \Big( \ln \tilde Z_N
    - \tilde \omega (H) + \tilde {\omega} (\tilde H) \Big)
\end{eqnarray}

\no which explicitly reads:
\begin{eqnarray}\label{nonprecise}
    p_N=\frac{1}{N}\ln Z_N  & \geqslant &  \ln 2 + \sum_{g=1}^p \a_g \ln \cosh \, r_g +
\\
  &&  
     +\frac{1}{2}  \sum_{g, k =1}^p   J_{g,k} \tanh r_g \tanh r_k
     + \sum_{g=1}^p h_g \tanh r_g
\\
  &&  - \sum_{g=1}^p \alpha_g \, r_g \, \tanh r_g +
          \nonumber\\
 &&  +  \frac{1}{2N}  \sum_{g=1}^p  \frac{J_{g,g}}{\a_g}  (\tanh r_g) ^2     +\frac{1}{2N}  \sum_{g=1}^p \frac{1}{\a_g J_{g,g}}
\\
\end{eqnarray}

Taking the lim inf over $N$ and the supremum in the variables $r_g$  the left hand side we get the (\ref{plow})
after performing the change of variables $\bar m_g=\tanh r_g$ .

\end{proof}

\subsection{Exact solution of the model}
	
Though the functions $p_{LOW}$ and $p_{UP}$ are different, it is easily checked that they share the same local suprema.
Indeed, if we differentiate both functions with respect to parameters $\bar m_g$, we see that the extremality
conditions are given in both cases by the Mean Field Equations:
\begin{equation}\label{MFE}
    \bar m_g  =  \tanh \Big ( \sum_{k=1}^p \frac{ J_{g, k} + J_{k, g}}{2 \, \a_g} \bar m_k   +  \frac{h_g}{\a_g} \Big ) \quad g=1..p
 \end{equation}

If we now use these equations to express $\tanh^{-1} m_i$ as a function of $m_i$ and we substitute back
into $p_{UP}$ and $p_{LOW}$ we get the same function:
%
%
\begin{equation}\label{presfunc}
p  = - \frac{1}{2}  \sum_{g, k =1}^p   J_{g,k} \bar m_g \bar m_k
         - \sum_{g=1}^p \a_g \frac{1}{2} \ln \frac{1- \bar m_g^2}{4}.
\end{equation}

Since this function returns the value of the pressure when the vector $(\bar m_1,.., \bar m_p)$ corresponds to an extremum, and this is the same both for $p_{LOW}$ and $p_{UP}$, we have proved the following:

\begin{theorem}
Given a hamiltonian as defined in (\ref{ham1}), and defining the pressure per particle as
$\displaystyle{p_N=\frac 1 N \ln Z}$,
given parameters $J_{i, j}$ and $h_i$, the thermodynamic limit
$$
\lim_{N\rightarrow \infty} p_N = p
$$
\no of the pressure exists,
and can be expressed in one of the following equivalent forms:
\begin{itemize}
		\item[a)] $\displaystyle{p = \sup_{\bar m_1,.., \bar m_p} \ p_{LOW}}$
		\item[b)] $\displaystyle{p = \sup_{\bar m_1,.., \bar m_p} \ p_{UP}}$
\end{itemize}
\end{theorem}

\section{An analytic result for a two-population model}\label{maxima}

The form we derived for the pressure can be rightfully considered a solution
of the statistical mechanical model, since it expresses the thermodynamic properties
of a large number of particles in terms of a finite number of parameters.

Nevertheless, the equations of state cannot be solved explicitly in terms of
the parameters: indeed, even the phase diagram for the two-population case has only
been characterised fully in a subset of our parameter space, in which it has been
found useful for a few physical applications \cite{cohen, kincaid, kulske}. This gives
us a feeling of how the mean field assumption, being simplistic from one point of view,
can given rise to models exhibiting non-trivial behaviour.

In this section we shall focus on the two-population case, which is the case considered
in the applications of the next chapter, and find an analytic result concerning the
maximum number of equilibrium states arising from our equations of state.
In particular we shall prove that, for any choice of the parameters, the
total number of local maxima for the function $p(\bar m_1, \bar m_2)$ is less or equal to five.

By applying a convenient relabelling to the model's parameters, we get the mean field equations for our
two-population model in the following form:
\begin{displaymath}
\left\{ \begin{array}{lll}
\bar m_1 & = & \tanh(J_{11}\alpha \bar m_1   + J_{12}(1-\alpha)\bar m_2 +  h_1 )
\\
\bar m_2 & =  & \tanh(J_{12}\alpha\bar m_1 + J_{22}(1-\alpha)\bar m_2    +  h_2 )
\end{array}, \right.
 \end{displaymath}

\no and correspond to the stationarity conditions of $p(\bar m_1, \bar m_2)$. So, a subset of
solutions to this system of equations are local maxima, and some among them correspond
to the thermodynamic equilibrium.


These equations give a two-dimensional generalization of the Curie-Weiss mean field equation.
Solutions of the classic Curie-Weiss model can be analysed by elementary geometry:
in our case, however, the geometry is that of 2 dimensional maps, and it pays to recall
that Henon's map, a simingly harmless 2 dimensional diffeomorhism of $\R^2$, is known to exhibit
full-fledged chaos. Therefore, the parametric dependence of solutions, and in particular the number
of solutions corresponding to local maxima of $p(\bar m_1, \bar m_2)$, is  in no way
apparent from the equations themselves.

We can, nevertheless, recover some geometric features from the analogy with one-dimensional picture.
For the classic Curie-Weiss equation, continuity and the Intermediate Value Theorem from elementary calculus
assure the existence of at least one solution. In higher dimensions we can resort to the analogous result,
Brouwer's Fixed Point Theorem, which states that any continuous map on a topological closed ball
has at least one fixed point.
This theorem, applied to the smooth map $R$ on the square $[-1,1]^2$, given by
\begin{displaymath}
\left\{ \begin{array}{lll}
R_1(\bar m_1,\ \bar m_2) & = & \tanh(J_{11}\alpha \bar m_1   + J_{12}(1-\alpha)\bar m_2 +  h_1 )
\\
R_2(\bar m_1,\ \bar m_2) & =  & \tanh(J_{12}\alpha\bar m_1 + J_{22}(1-\alpha)\bar m_2    +  h_2 )
\end{array} \right.
 \end{displaymath}
establishes the existence of at least one point of thermodynamic equilibrium.

We can gain further information by considering the precise form of the equations:
by inverting the
hyperbolic tangent in the first equation,  we can $\bar m_1$ as a function of $\bar m_2$, and vice-versa for
the second equation.
Therefore, when $J_{12}\neq0$ we can rewrite the equations in the following fashion:
\begin{eqnarray} \label{gamma1}
\left\{ \begin{array}{lll}
\bar m_2 & = & \displaystyle{ \frac{1}{J_{12}(1-\alpha)} (\tanh^{-1} \bar m_1 -  J_{11}\alpha \bar m_1- h_1 ) }
\\
\bar m_1 & = & \displaystyle{\frac{1}{J_{12}\alpha} (\tanh^{-1}\bar m_2 -  J_{22}(1-\alpha) \bar m_2- h_2 ) }
\end{array} \right.
 \end{eqnarray}

Consider, for example, the first equation: this defines a function $\bar m_2(\bar m_1)$, and we shall
call its graph {\it curve $\gamma_1$}. Let's
consider the second derivative of this function:
$$
\frac{\partial^2\bar m_2}{\partial \bar m_1 ^2}=-\frac{1}{J_{12}(1-\alpha)} \cdot \frac{2 \bar m_1}{(1-\bar m_1^2)^2}.
$$

\begin{figure}
    \centering
    \includegraphics[width=8 cm]{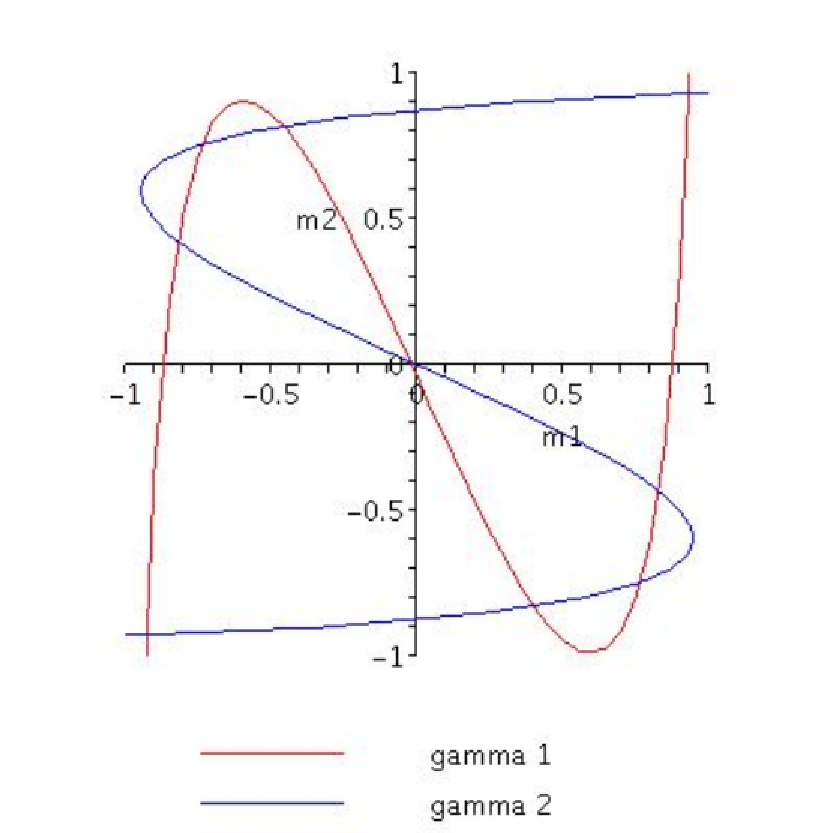}
    \caption{The crossing points correspond to solutions of the mean field equations}
    \label{9points}
\end{figure}

We see immediately that this second derivative is strictly increasing,
and that it changes sign exactly at zero.
This implies that $\g_1$ can be divided into three monotonic pieces, each having
strictly positive {\it third} derivative as a function of $\bar m_1$.
The same thing holds for the second equation, which defines a function $\bar m_1(\bar m_2)$,
and a corresponding {\it curve $\gamma_2$}.
An analytical argument easily establishes that there exist at most $9$
crossing points of $\g_1$ and $\g_2$  (for convenience we shall label the three monotonic pieces
of $\g_1$ as $I$, $II$ and $III$, from left to right): since $\g_2$, too, has a strictly positive third derivative,
it follows that it intersects each of the three monotonic pieces of $\g_1$ at most three times, and this
leaves
the number of intersections between $\g_1$ and $\g_2$ bounded above by 9 (see an example
of this in Figure \ref{9points}).

By definition of the mean field equations, the stationary points of the pressure correspond to
crossing points of $\gamma_1$ and $\gamma_2$.
Furthermore, common sense tells us that not all of these stationary points can be local maxima.
This is indeed true, and it is proved by the following:

\begin{proposition}\label{5max}
The function $p(\bar m_1, \bar m_2)$ admits at most 5 maxima.
\end{proposition}

To prove \ref{5max} we shall need the following:

\begin{lemma}\label{adjpoints}
Say $P_1$ and $P_2$ are two crossing points linked by a monotonic piece of one of the two functions
considered above. Then at most one of them is a local maximum of the pressure $p(\bar m_1, \bar m_2)$.
\end{lemma}

\begin{proofof}{Lemma \ref{adjpoints}}
The proof consists of a simple observation about the meaning of our curves.
The mean field equations as stationarity conditions for the pressure,
so each of $\gamma_1$ and $\gamma_2$
are made of points where one of the two components of the gradient
of $p(\bar m_1, \bar m_2)$ vanishes. Without loss of generality assume that $P_1$ is a maximum,
and that the component that vanishes on the piece of curve that links $P_1$ to $P_2$ is
$\ds{\frac{\partial p}{\partial \bar m_1}}$.

Since $P_1$ is a local maximum, $p(\bar m_1, \bar m_2)$ locally increases on the piece of curve $\gamma$.
On the other hand, the directional derivative of $p(\bar m_1, \bar m_2)$ along $\gamma$  is given by
$$
\mathbf{\hat t} \cdot \nabla p
$$

\no where $\mathbf{\hat t}$ is the unit tangent to $\gamma$. Now we just need to notice that
by assumptions for any
point in $\gamma$ $\mathbf{\hat t}$ lies in the same quadrant, while $\nabla p$ is vertical with
a definite verse.
This implies that the scalar product giving directional derivative is strictly non-negative over all $\gamma$,
which prevents $P_2$ form being a maximum.
\end{proofof}

\begin{proofof}{Proposition \ref{5max}}
The proof considers two separate cases:

\begin{itemize}
\item[a)] All crossing points can be joined in a chain by using monotonic pieces of curve such as the one
          defined in the lemma;
\item[b)] At least one crossing point is linked to the others only by non-monotonic pieces of curve.
\end{itemize}

In case a), all stationary can be joined in chain in which no two local maxima can be nearest neighbours,
by the lemma.
Since there are at most 9 stationary points, there can be at most 5 local maxima.

For case b) assume that there is a point, call it $P$, which is not linked to any
other point by a monotonic piece of curve.
Without loss of generality, say that $P$ lies on $I$ (which, we recall, is defined as the leftmost
monotonic piece of $\gamma_1$). By assumption, $I$ cannot contain other crossing points
apart from $P$, for otherwise $P$ would be monotonically linked to at least one of them, contradicting
the assumption.
On the other hand,  each of $II$ and $III$  contain at most $3$ stationary points, and, by Lemma \ref{adjpoints}, at most $2$ of these are maxima. So we have at most $2$ maxima on each of
$II$ and $III$, and and at most 1 maximum on $I$,  which leaves the total bounded above by $5$.
The cases in which $P$ lies on $II$, or on $III$, are proved analogously, giving the result.

\end{proofof}

%% file: capitolo_case_studies.tex
\resettheoremcounters

\chapter{Case studies}

In previous chapters we defined a model which, generalizing well known tools from econometrics, provides a viable
approach to study phenomena of human interaction. Its well-posedness as an equilibrium statistical mechanical model,
proved in the last chapter, though supporting the idea that modelling social phenomena working from the bottom up\footnote{
that is, starting from individual interactions and trying to establish patterns that might be at work on a larger
scale} may be feasible, doesn't imply the relevance of the proposed tool to any actual scenario: indeed, for 
any model such relevance may
only be established as a result of success in describing, and most importantly predicting events from the real world.

There are many possible instances from the social sciences to which quantitative modelling is an appealing prospective.
Due to the increasingly global nature of human mobility, one particularly timely social issue is immigration. The applicability
of our model to immigration matters was considered in References \cite{cgm} and \cite{cgg}. Reference \cite{cgg} analyses
how the microscopic assumptions of the model reflect the tendency of individuals to act consistently with their cultural legacy
as well as with what they identify as their social group, which are both tenets in the field of social psychology.
The numerical analysis carried out in Reference \cite{cgm} shows how such simple assumptions are enough for the model
to identify regimes in which a global change in a cultural trait is triggered by a small fraction of immigrants
interacting with a large population of residents.

The descriptive power shown by the model in the case of immigration further supports the view that equilibrium statistical
mechanics can play a role in a quantitative theory of social phenomena. However, though qualitatively inspiring, the
immigration scenario seems ill-suited as a first quantitative case study, due to the intrinsic difficulty of finding
a database that characterizes such a social issue adequately. We therefore turn to the problem of giving our model a
first implementation on some ``simpler'' matters.

The aim of this chapter is two-fold. On one hand we are interested in assessing the simplest instance
of the model considered in the last chapter, that is a mean field model where the population has been
partitioned into two groups, based on their geographical residence, so that the model generalizes a discrete choice
model with one binary attribute.

On the other hand, we'd like to propose two simple procedures of model estimation, that we feel 
might be very appealing for models at an early stage of development. The first procedure is statistical in nature, and
it's based on a method developed by Berkson \cite{berkson}, whereas the second takes a statistical mechanical perspective
by considering the role played by the fluctuations of the main observable quantities for the model.

\section{The model}

We consider a population of individuals facing with a ``YES/NO''
question, such as choosing between marrying through a religious or
a civil ritual, or voting in favor or against of death penalty in
a referendum. We index individuals by $i, \ i=1...N$, and assign a
numerical value to each individual's choice $\s_i$ in the following
way:
$$
\s_i = \left\{ \begin{array}{lll}
     +1 \textrm{ if $i$ says YES}
    \\
     -1 \textrm{ if $i$ says NO}
    \end{array}, \right.
$$
Consistently with the many population Curie-Weiss model analysed in the last chapter,
which as we saw generalises the multinomial logit model described in chapter \ref{ds},
we assume that the joint probability distribution of these choices is well approximated by
a Boltzmann-Gibbs distribution corresponding to the following Hamiltonian
$$
H_N(\s)=-\sum_{i,l=1}^N J_{il}\s_i\s_l - \sum_{i=1}^N h_i \s_i.
$$

Heuristically, this distribution favours the agreement of
people's choices $\s_i$ with some external influence $h_i$ which
varies from person to person, and at the same time favours
agreement of a couple of people whenever their interaction
coefficient $J_{il}$ is positive, whereas favors disagreement
whenever $J_{il}$ is negative.

Given the setting, the model consists of two basic steps:
\begin{itemize}
    \item[1)] A parametrization of quantities $J_{il}$ and of $h_i$,
    \item[2)] A systematic procedure allowing us to ``measure'' the parameters characterizing the model, starting
              from statistical data (such as surveys, polls, etc).
\end{itemize}

The parametrization must be chosen to fit as well as possible the
data format available, in order to define a model which is able to make
good use of the increasing wealth of data available through
information technologies.

\section{Discrete choice}

Let us first consider our model when it ignores interactions
$J_{il}\; \equiv \; 0 \; \forall \; i,l \; \in \; (1,...,N)$, that is
$$
H_N(\s)= - \sum_{i=1}^N h_i \s_i.
$$

The model shall be applied to data coming from surveys, polls, and censuses, which means that
together with the answer to our binary question, we shall have access to information characterizing
individuals from a socio-economical point of view.
We can formalize such further information by assigning to each person a vector of \emph{socio-economic attributes}
$$
a_i=\{a_i^{(1)}, a_i^{(2)},..., a_i^{(k)}\}
$$
where, for instance,
$$
a_i^{(1)} = \left\{ \begin{array}{l}
    1 \textrm{ for $i$ Male}
    \\
    0 \textrm{ for $i$ Female}
    \end{array}, \right.
$$
and
$$
a_i^{(2)} = \left\{ \begin{array}{l}
    1 \textrm{ for $i$ Employee}
    \\
    0 \textrm{ for $i$ Self-employed}
    \end{array}, \right.
$$
etc.

As we have seen in chapter \ref{ds}, the general setting of the multinomial logit
allows to exploit the supplementary data by assuming that
$h_i$ (which is the ``field'' influencing the choice of $i$) is a function
of the vector of attributes $a_i$. Since for the sake of simplicity we choose
our attributes to be binary variables, so that the most general form for $h_i$ turns
out to be linear
$$
h_i=\sum_{j=1}^k \a_j a_i^{(j)}+\a_0
$$
and the model's parameters are given by the components of the vector $\a=\{ \a_0, \a_1,..., \a_k \}$.
It's worth pointing out that the parameters $\a_j$, $j=0...k$  do not depend on the specific individual $i$.

We know that discrete choice theory holds that, when making a choice, each
person weights out various factors such as his own gender, age, income,
etc, as to maximize in probability the benefit arising from
his/her decision. Parameters $\a$ tell us the relative weight
(i.e. their relative importance importance) that the various socio-economic factors have
when people are making a decision with respect to our binary
question.
The parameter $\a_0$ does not multiply any specific attribute, and thus
it is a homogeneous influence which is felt by all people
in the same way, regardless of their individual characteristics.
A discrete choice model is considered good when the parametrized
attributes are very suitable for the specific choice, so that
the parameter $\a_0$ is found to be small in comparison to the
attribute-specific ones.

We have shown in chapter \ref{ds} that elementary statistical mechanics
gives us the probability of
an individual $i$ with attributes $a_i$ answering ``YES'' to our
question as:
\begin{eqnarray*}
p_i&=&P(\s_i=1)=\frac{e^{h_i}}{e^{h_i}+e^{-h_i}},
\\
h_i&=&\sum_{j=1}^k \a_j a_i^{(j)}+\a_0,
\end{eqnarray*}
which as we saw is equivalent to the result obtained by applying
economics' utility maximization principle to a random utility
with Gumbel disturbances.
Therefore collecting the choices made by a relevant number of
people, and keeping track of their socio-economic attributes,
allows us to use statistics in order to find the value of $\a$
for which our distribution best fits the real data. This in turn
allows to assess the implications on aggregate behavior if we
apply incentives to the population which affect specific
attribute, as can be commodity prices in a market situation.
%

\section{Interaction}

The kind of model described in the last section has been
successfully used by econometrics for the last thirty years
\cite{mcfadden}, and has opened the way to the quantitative study
of social phenomena. Such models, however, only apply to
situations where the functional relation between the people's
attributes $\a$ and the population's behavior is a smooth one: it is
ever more evident, on the other hand, that behavior at a societal
level can be marked by sudden jumps \cite{bouchaud2, salganik,
kuran}.

There exist many examples from linguistics, economics, and sociology
where it has been observed how the global behaviour of
large groups of people can change in an abrupt manner as a
consequence of slight variations in the social structure (such as,
for instance, a change in the pronunciation of a language due to a
little immigration rate, or as a substantial decrease in crime rates
due to seemingly minor action taken by the authorities)
\cite{critmass, gladwell, kuran}. From a statistical mechanical
point of view, these abrupt transitions may be considered
as phase transitions caused by the interaction between
individuals, and this is what led us to consider in this thesis
the interesting mapping between discrete choice econometrics and
the Curie-Weiss theory, first stated in \cite{durlauf}.

We then go back to studying the general interacting model
\begin{equation}\label{ham}
H_N(\s)=-\sum_{i,l=1}^N J_{il}\s_i\s_l - \sum_{i=1}^N h_i \s_i,
\end{equation}
while keeping
$$
h_i=\sum_{j=1}^k \a_j a_i^{(j)}+\a_0.
$$

We now need to find a suitable parametrization for the interaction coefficients $J_{il}$.
Since each person is characterized by $k$ binary socio-economic attributes,
the population can be naturally partitioned into $2^k$ subgroups, so that using the mean-field
assumptions allows one to
rewrite the model in terms of subgroup-specific magnetizations $m_g$, as in the general
Hamiltonian (\ref{generalham}). Equation (\ref{generalham}) is general enough to consider
populations with different relative sizes (such as one in which residents make up a much larger
share of population than immigrants): nevertheless, it turns out that the mean-field assumption
implies a relation of direct proportionality between interaction coefficients and population
sizes, that might be considered innatural.

The approach taken in this thesis, therefore, is to consider sub-populations of comparable
size, and model them in the thermodynamic limit as having equal size. In specific, in all cases
we divide the data into two geographical regions which have a similar population.
This ``equal size'' assumption can be considered as part of the modelling process:
by using it to analyze data, as we do here, we can gain insights on how to relax it in future refinements
of the model.
So, for the time being, let $J_{il}$ depend explicitly on a partition of
sub-populations of equal sizes. By using the mean-field assumption we can express this as follows
\begin{eqnarray*}
J_{il}=\frac{1}{2^k \, N}J_{gg'}, \ \textrm{if $i \in g$ and $l \in g'$},
\end{eqnarray*}
where $g$ and $g'$ are two sub-population (not necessarily distinct). This in turn allows us to rewrite (\ref{ham}) as
\begin{eqnarray*}\label{intens}
H_N(\s)=-\frac{N}{2^k}(\sum_{g,g'=1}^{2^k} J_{gg'}m_g m_{g'} +
\sum_{g=1}^{2^k} h_g m_g)
\end{eqnarray*}
where $m_g$ is the average opinion of group $g$:
$$
m_g=\frac{1}{2^k \, N}\sum_{i=(g-1)N/2^k+1}^{g \, N/2^k} \s_i.
$$

We readily see how this is the many-population model considered in the previous chapter, and this gives us a solid
microscopic foundation for the theory. Indeed, the results we obtained through relatively elementary mathematics
establish rigourously the existence of the model's thermodynamic limit, as well as its factorization properties,
and just as importantly provide us with a closed form for the thermodynamic state equations.

Therefore if we are willing to test how well the model's assumptions compare with real data, we can use
these equations as the main tool for a procedure of statistical estimation.
Here we shall confront the simple case where $k=1$. This is a bipartite model which, as we know from the last
chapter, can have at most five metastable
equilibrium states, given by the thermodynamically stable solutions to the following equations:
\begin{eqnarray}\label{mfe}
\bar m_1 &=& \tanh(J_{11} \bar m_1 + J_{12} \bar m_2 + h_1)
\\
\bar m_2 &=& \tanh(J_{21} \bar m_1 + J_{22} \bar m_2 + h_2)
\end{eqnarray}

Equation (\ref{MFE}) which was derived from the model's exact solution shows that the
equilibrium state equations for a system consisting of two parts of equal size do not carry two
different parameters $J_{12}$ and $J_{21}$, but that, even if these two parameters were different
in the Hamiltonian, what characterizes each of the two subparts is rather their average
$(J_{12}+J_{21})/2$. 
We keep $J_{12}$ and $J_{21}$ as two distinct parameters
throughout the statistical application in order to use them as a consistency test: we shall be
able to consider systems to be in equilibrium only if $J_{12}-J_{21}=0$.

The state equations (\ref{mfe}) allow us, in particular, to write the probability of $i$ choosing YES in a closed form,
similar to the non-interacting one:
\begin{equation}\label{prob}
p_i=P(\s_i=1)=\frac{e^{U_g}}{e^{U_g}+e^{-U_g}},
\end{equation}
where
$$
U_g=\sum_{g'=1}^2 J_{g,g'} \bar m_{g'} + h_g.
$$

This is the basic tool needed to estimate the model starting from real data. We describe
the estimation procedure in the next section.

\section{Estimation}

We have seen that according to the model an individual $i$ belonging to group $g$ has
probability of choosing ``YES'' equal to
$$
p_i=\frac{e^{U_g}}{e^{U_g}+e^{-U_g}}
$$
where
$$
U_g=\sum_{g'} J_{g,g'} \bar m_{g'} + h_g.
$$

The standard approach of statistical estimation for discrete
models is to maximize the probability of observing a sample of
data with respect to the parameters of the model (see
e.g. \cite{benakiva}). This is done by maximizing the likelihood
function
$$
L=\prod_i p_i
$$
with respect to the model's parameters, which in our case consist of the interaction
matrix $J$ and the vector $\alpha$.

Our model, however, is such that $p_i$ is a function of the equilibrium states $m_g$,
which in turn are {\it discontinuous} functions of the model's parameters.
This problem takes away much of the appeal of the maximum likelihood procedure,
and calls for a more feasible alternative.

The natural alternative to maximum likelihood for problems of
model regression is given by the least squares method
\cite{greene}, which simply minimizes the squared norm of the
difference between observed quantities, and the model's
prediction. Since in our case the observed quantities are the
empirical average opinions $\tm_g$, we need to find the parameter values
which minimize
\begin{equation}\label{least}
    \sum_g (\tm_g - \tanh U_g)^2,
\end{equation}
which in our case correspond to satisfying as closely as possible
the state equations (\ref{mfe}) in squared norm. This, however, is
still computationally cumbersome due to the non-linearity of the
function $\tanh(U_j)$. This problem has already been encountered
by  Berkson back in the nineteen-fifties, when developing a
statistical methodology for bioassay \cite{berkson}: this is an
interesting point, since this stimulus-response kind of experiment
bears a close analogy  to the natural kind of
applications for a model of social behavior, such as linking
stimula given by incentive through policy and media, to behavioral
responses on part of a population. 

The key observation in Berkson's paper is that, since $U_g$ is a linear function of the model's parameters,
and the function $\tanh(x)$ is invertible, a viable modification to least squares is given by minimizing the
following quantity, instead:
\begin{equation}\label{berk}
    \sum_g (\arctanh \tm_g -  U_g)^2.
\end{equation}
This reduces the problem to a linear least squares problem which can be handled with standard statistical
software, and Berkson finds an excellent numerical agreement between this method and the standard least squares procedure.

There are nevertheless a number of issues with Berkson's approach,
which are analyzed in \cite{benakiva}, pag. 96. All the problems
arising can be traced to the fact that to build (\ref{berk}), we
are collecting the individual observations into subgroups, each of
average opinion $m_g$. The problem is well exemplified by the case
in which a subgroup has average opinion $m_g \equiv \pm 1$: in
this case $\arctanh m_g=-\infty$, and the method breaks down.
However the event $m_g \equiv \pm 1$ has a vanishing probability
when the size of the groups increases, so that the method behaves
properly for large enough samples.
\newline
The proposed measurement technique is best elucidated by showing a
few simple concrete examples, which we do in the next section.

\section{Case studies}

We shall carry out the estimation program for real situations which
correspond to a very simple case of our model.
The data was obtained from periodical censuses carried out by Istat\footnote{Italian National Institute of Statistics}:
since census data concerns events which are recorded in official documents, for a large number of people, we find it
to be an ideal testing ground for our model.

For the sake
of simplicity, individuals  are described  by a single binary attribute characterizing their place of residence
(either Northern or Southern Italy) and we chose, among the several possible case studies, the ones for which 
choices are likely to involve peer interaction in a  major way.

The first phenomenon we choose to study concerns the share of people who
chose to marry through a religious ritual, rather than through a
civil one. The second case deals with divorces:  here individuals are faced with the choice
 of a consensual/ non-consensual divorce.
The last test we perform regards the study of suicidal tendencies, in particular the mode of execution.
\newline

\subsection{Civil vs religious marriage in Italy, 2000-2006}

To address this first  task  we use data from the annual report on
the institution of marriage compiled by Istat in the seven years going from
$2000$ to $2006$.
The reason for choosing this specific social question is both a methodological and a conceptual one.

Firstly, we are motivated by the exceptional quality of the data
available in this case, since it is a census which concerns a
population of more  than $250$ thousand people per year, for seven
years. This allows us some leeway from the possible issues
regarding the sample size, such as the one highlighted in the last
section. And just as importantly the availability of a time series
of data measured at even times also allows to check the
consistency of the data as well as the stability of the
phenomenon.

Secondly, marriage is probably one of the few matters where a great number of individuals
make a genuine choice concerning their life that gets recorded in an official document, as opposed
to what happens, for example, in the case of opinion polls.

We choose to study the data with one of the simplest forms of the model: individuals are divided
according to only to a binary attribute $a^{(1)}$, which takes value $1$ for people from Northern
Italy, and $0$ for people form Southern Italy.
In the formalism of Section $2$, therefore, the model is defined
by the Hamiltonian
\begin{eqnarray*}
H_N(\s) &=& - \frac{N}{2} ( J_{11} m_1^2 + (J_{12}+J_{21}) m_1 m_2 + J_{22} m_2^2 +  h_1 m_1 +  h_2 m_2),
\\
h_i     &=& \a_1 a_i^{(1)}+\a_0,
\end{eqnarray*}
and the state equations to be used for Berkson's statistical procedure are given by (\ref{prob}).

\begin{table}
\begin{center}
{\begin{tabular}{@{}cccccccc@{}}
  & \multicolumn{7}{c}{\textbf{ \% of religious marriages, by year}} \\
\midrule
\textbf{Region} & \textbf{2000} & \textbf{2001} & \textbf{2002} & \textbf{2003} &
\textbf{2004} & \textbf{2005} & \textbf{2006} \\ \midrule
\textbf{Northern Italy} & 68.35 & 64.98 & 61.97 & 60.90 & 57.91 & 55.95 & 54.64 \\
\textbf{Southern Italy} & 81.83 & 80.08 & 79.32 & 79.02 & 76.81 & 76.52 & 75.46 \\
\bottomrule
\end{tabular}}
\caption{Percentage of religious marriages, by year and geographical region}\label{ts1}
\end{center}
\end{table}

Table \ref{ts1} shows the time evolution of the share of men
choosing to marry through a religious ritual: the population is
divided in two geographical classes. The first thing worth noticing is that
these shares show a remarkable stability over the seven-year
period: this confirms how, though arising  from choices made by
distinct individuals, who bear extremely different personal
motivations, the aggregate behavior can be seen as an observable
feature characterizing society as a whole.


In order to apply Berkson's method of estimation, we choose gather
the data into periods of four years, starting with $2000-2003$,
then $2001-2004$, etc. Now, if we label the share of men in group
$g$ choosing the religious ritual in a specific year (say in
$2000$) by $m_g^{2000}$, we have that the quantity that ought to
be minimized in order to estimate the model's parameters for the
first period is the following, which we label $X^2$:
\begin{eqnarray*}
   X^2 &=& \sum_{year=2000}^{2003} \sum_{g=1}^2 (\arctanh m^{year}_g -  U_g^{year})^2,
\\
    U_g^{year}&=&\sum_{g'=1} ^2J_{g,g'} m_{g'}^{year} + h_g,
\\
    h_g     &=& \a_1 a_g^{(1)}+\a_0.
\end{eqnarray*}

The results of the estimation for the four periods are shown in Table \ref{J1}, whereas Table \ref{H1} shows the corresponding
estimation for a discrete choice model which doesn't take into account interaction.
\begin{table}
\begin{center}
{\begin{tabular}{@{}ccccc@{}}
 & \multicolumn{4}{c}{ \textbf{4-year period} } \\
\midrule
\textbf{Parameter} & \textbf{2000-2003} & \textbf{2001-2004} & \textbf{2002-2005} & \textbf{2003-2006} \\
\midrule
{$\a_0$} & -0.10 $\pm$ 0.42 & -0.16 $\pm$ 0.15 & -0.18 $\pm$ 0.10 & -0.13 $\pm$ 0.01 \\
{$\a_1$} & 0.20 $\pm$ 0.59 & 0.20 $\pm$ 0.22 & 0.16 $\pm$ 0.14 & 0.14 $\pm$ 0.01 \\
{$J_1$} & 1.16 $\pm$ 0.41 & 1.09 $\pm$ 0.16 & 1.01 $\pm$ 0.11 & 1.02 $\pm$ 0.01 \\
{$J_2$} & 1.29 $\pm$ 0.89 & 1.40 $\pm$ 0.33 & 1.45 $\pm$ 0.21 & 1.36 $\pm$ 0.01 \\
{$J_{12}$} & -0.21 $\pm$ 0.89 & -0.10 $\pm$ 0.33 & 0.03 $\pm$ 0.21 & -0.01 $\pm$ 0.01 \\
{$J_{21}$} & 0.09 $\pm$ 0.41 & 0.02 $\pm$ 0.16 & -0.01 $\pm$ 0.11 & 0.01 $\pm$ 0.01 \\
\bottomrule
\end{tabular}}
\caption{Religious vs civil marriages: estimation of the interacting model}\label{J1}
\end{center}
\end{table}

\begin{table}
\begin{center}
{\begin{tabular}{@{}ccccc@{}}
 & \multicolumn{4}{c}{ \textbf{4-year period} } \\
\midrule
\textbf{Parameter} & \textbf{2000-2003} & \textbf{2001-2004} & \textbf{2002-2005} & \textbf{2003-2006} \\
\midrule
{$\a_0$} & 0.67 $\pm$ 0.15 & 0.63 $\pm$ 0.03 & 0.61 $\pm$ 0.06 & 0.58 $\pm$ 0.03 \\
{$\a_1$} & -0.41 $\pm$ 0.1 & -0.43 $\pm$ 0.04 & -0.45 $\pm$ 0.08 & -0.46 $\pm$ 0.04 \\
\bottomrule
\end{tabular}}
\caption{Religious vs civil marriages: estimation of the non-interacting model}\label{H1}
\end{center}
\end{table}

\subsection{Divorces in Italy, 2000-2005}

The second case study uses data from the annual report compiled by Istat in the six years
going from $2000$ to $2005$. The data show how divorcing couples chose between a consensual and a non-consensual
 divorce in Northern  and Southern Italy.
As shown in Table \ref{ts2} here too, when looking at the ratio among consensual versus the total divorces, the data show a remarkable stability.

Again we gather the data into periods of four years and Table \ref{J2} presents the estimation of our model's parameters for the whole 
available period, while in Table \ref{H2} we show the corresponding fit by the non-interacting discrete choice model.

We notice that the  estimated parameters have some analogies with the preceding case study in that here too the cross 
interactions $J_{12}, \ J_{21}$ are statistically close to zero whereas the diagonal values $J_{11}, \ J_{22}$ are both greater 
than one suggesting an interaction scenario characterized by multiple equilibria \cite{gaco}. Furthermore, in both cases the 
attribute-specific parameter $\a_1$
is larger than the generic parameter $\a_0$ in the interacting model (Tables 2 and 5), as opposed to what we see in the
non-interacting case
(Tables 3 and 6): this suggests that by accounting for interaction we might be able to better evaluate the role played by
socio-economic attributes.

\begin{table}
\begin{center}
{\begin{tabular}{@{}ccccccc@{}}
  & \multicolumn{6}{c}{\textbf{ \% of consensual divorces, by year}} \\
\midrule
\textbf{Region} & \textbf{2000} & \textbf{2001} & \textbf{2002} & \textbf{2003} &
\textbf{2004} & \textbf{2005} \\
\midrule
{\textbf Northern Italy} & 75.06 & 80.75 & 81.32 & 81.62 & 81.55 & 81.58 \\
{\textbf Southern Italy} & 58.83 & 72.80 & 71.80 & 72.61 & 72.76 & 72.08 \\
\bottomrule
\end{tabular}}
\caption{Percentage of consensual divorces, by year and geographical region}\label{ts2}
\end{center}
\end{table}

\begin{table}
\begin{center}
{\begin{tabular}{@{}cccc@{}}
 & \multicolumn{3}{c}{ \textbf{4-year period} } \\
\midrule
\textbf{Parameter} & \textbf{2000-2003} & \textbf{2001-2004} & \textbf{2002-2005}  \\
\midrule
{$\a_0$} & 0.02 $\pm$ 0.06 & -0.08 $\pm$ 0.01 & -0.07 $\pm$ 0.01 \\
{$\a_1$} & -0.25 $\pm$ 0.08 & -0.22 $\pm$ 0.01 & -0.23 $\pm$ 0.01 \\
{$J_1$} & 1.59 $\pm$ 0.14 & 1.64 $\pm$ 0.01 & 1.66 $\pm$ 0.01 \\
{$J_2$} & 1.16 $\pm$ 0.06 & 1.25 $\pm$ 0.01 & 1.25 $\pm$ 0.01 \\
{$J_{12}$} & -0.05 $\pm$ 0.06 & 0.01 $\pm$ 0.01 & 0.00 $\pm$ 0.01 \\
{$J_{21}$} & -0.08 $\pm$ 0.14 & 0.00 $\pm$ 0.01 & -0.01 $\pm$ 0.01 \\
\bottomrule
\end{tabular}}
\caption{Consensual vs non-consensual divorces: estimation of the interacting model}\label{J2}
\end{center}

\end{table}
\begin{table}
\begin{center}
{\begin{tabular}{@{}cccc@{}}
 & \multicolumn{3}{c}{ \textbf{4-year period} } \\
\midrule
\textbf{Parameter} & \textbf{2000-2003} & \textbf{2001-2004} & \textbf{2002-2005}  \\
\midrule
{$\a_0$} & 0.41 $\pm$ 0.13 & 0.48 $\pm$ 0.01 & 0.480046 $\pm$ 0.01 \\
{$\a_1$} & 0.28 $\pm$ 0.18 & 0.25 $\pm$ 0.02 & 0.261956 $\pm$ 0.01 \\
\bottomrule
\end{tabular}}
\caption{Consensual vs non-consensual divorces: estimation of the non-interacting model}\label{H2}
\end{center}
\end{table}

\subsection{Suicidal tendencies in Italy, 2000-2007}

The last case study deals with suicidal tendencies in Italy, again following the  annual report compiled by Istat in the
eight years  from $2000$ to $2007$, and we use the same geographical attribute used for the former two studies.

The data in Table \ref{ts3} shows the percentage of deaths due to hanging as a mode of execution. The topic of suicide is
of particular relevance to sociology: indeed, the very first systematic quantitative treatise in the social sciences was
carried out by \'Emile Durkheim \cite{durkheim}, a founding father of the subject, who was puzzled by how a phenomenon as
unnatural as suicide could arise with the astonishing regularity that he found. Such a regularity as even been dimmed the
``sociology's one law'' \cite{pope}, and there is hope that the connection to statistical mechanics might eventually shed
light on the origin of such a law.

Mirroring the two previous case studies, we present the time series in Table \ref{ts3}, whereas Table \ref{J3} shows the
estimation results for the interacting model, and Table \ref{H3} are the estimation results for the discrete choice model.
Again, the data agrees with the analogies found for the two previous case studies. \newline

\begin{table}
\begin{center}
{\begin{tabular}{@{}ccccccccc@{}}
  & \multicolumn{8}{c}{\textbf{ \% suicides by hanging}} \\
\midrule
\textbf{Region} & \textbf{2000} & \textbf{2001} & \textbf{2002} & \textbf{2003} &
\textbf{2004} & \textbf{2005} & \textbf{2006} & \textbf{2007} \\
\midrule
\textbf{Northern Italy} & 34.17 & 37.02 & 35.83 & 34.58 & 35.21 & 36.23 & 33.57 & 38.08 \\
\textbf{Southern Italy} & 37.10 & 37.40 & 37.34 & 38.54 & 34.71 & 38.90 & 40.63 & 36.66 \\
\bottomrule
\end{tabular}}
\caption{Percentage of suicides with hanging as mode of execution, by year and geographical region}\label{ts3}
\end{center}
\end{table}

\begin{table}
\begin{center}
{\begin{tabular}{@{}cccccc@{}}
 & \multicolumn{5}{c}{ \textbf{4-year period} } \\
\midrule
\textbf{Parameter} & \textbf{2000-2003} & \textbf{2001-2004} & \textbf{2002-2005} & \textbf{2003-2006} & \textbf{2004-2007} \\
\midrule
{$\a_0$} & 0.01  $\pm$  0 & 0.02  $\pm$  0.01 & 0.01 $\pm$  0.01 & 0.02  $\pm$  0.01 & 0.02  $\pm$ 0.01 \\
{$\a_1$} & 0.01  $\pm$  0.01 & 0.00  $\pm$  0.01 & 0.00  $\pm$  0.01 & 0.00  $\pm$  0.01 & 0.00  $\pm$  0.01 \\
{$J_1$} & 1.09  $\pm$ 0.01 & 1.09  $\pm$  0.01 & 1.09  $\pm$  0.02 & 1.10  $\pm$  0.03 & 1.09  $\pm$  0.01 \\
{$J_2$} & 1.06  $\pm$  0.01 & 1.08  $\pm$ 0.01 & 1.08 $\pm$ 0.01 & 1.07 $\pm$  0.01 & 1.07  $\pm$  0.01 \\
{$J_{12}$} & 0  $\pm$ 0.01 & 0.00  $\pm$  0.01 & 0.00  $\pm$  0.01 & 0.00  $\pm$ 0.01 & 0.00  $\pm$  0.01 \\
{$J_{21}$} & 0  $\pm$  0.01 & 0.01  $\pm$  0.01 & 0.00  $\pm$  0.02 & 0.01  $\pm$ 0.03 & 0.01  $\pm$  0.01 \\
\bottomrule
\end{tabular}}
\caption{Suicidal tendencies: estimation of the interacting model}\label{J3}
\end{center}
\end{table}

\begin{table}
\begin{center}
{\begin{tabular}{@{}cccccc@{}}
 & \multicolumn{5}{c}{ \textbf{4-year period} } \\
\midrule
\textbf{Param.} & \textbf{2000-2003} & \textbf{2001-2004} & \textbf{2002-2005} & \textbf{2003-2006} & \textbf{2004-2007} \\
\midrule
{$\a_0$} & -0.25  $\pm$  0.02 & -0.27 $\pm$  0.03 & -0.26  $\pm$  0.03 & -0.24 $\pm$  0.04 & -0.25 $\pm$  0.05 \\
{$\a_1$} & -0.05  $\pm$  0.03 & -0.03  $\pm$ 0.04 & -0.04  $\pm$  0.04 & -0.07 $\pm$  0.06 & -0.04 $\pm$  0.07 \\
\bottomrule
\end{tabular}}
\caption{Suicidal tendencies: estimation of the non-interacting model}\label{H3}
\end{center}
\end{table}

\section{A statistical mechanical approach to model estimation}

We shall now estimate our model parameters using a different approach, which makes explicit use of the time
fluctuations of our main observable quantities $\tilde m_i$. This approach is not econometric, but typically
statistical mechanical, in that it equates fluctuations observed over time with fluctuations of a system
which is in an equilibrium which is defined by an {\it ensemble} of states rather than by
a single state. The problem of retracing a model's parameters from observable quantities in this context
 has been referred to in the literature as the ``inverse Ising problem'' (see e.g. \cite{monasson}).

We start from the usual model
\begin{eqnarray}\label{model}
H_N(\s) &=& - \frac{N}{2} ( J_{11} m_1^2 +  (J_{12} + J_{21}) m_1 m_2 + J_{22} m_2^2 +  h_1 m_1 +  h_2 m_2),
\\
\nonumber
h_i     &=& \a_1 a_i^{(1)}+\a_0,
\end{eqnarray}
and we shall analyze the data from our three case studies again using the model's state equations
\begin{eqnarray}\label{law}
\bar m_1 &=& \tanh(J_{11} \bar m_1 + J_{12} \bar m_2 + h_1),
\nonumber
\\
\bar m_2 &=& \tanh(J_{21} \bar m_1 + J_{22} \bar m_2 + h_2),
\end{eqnarray}
which, as we shall see, will now also provide us with the system's fluctuations as well as the average quantities.
Just as in the last section, we choose to use two distinct parameters $J_{12}$ and $J_{21}$ inside the state equations
(\ref{law})
instead of their average
 $\frac{1}{2}(J_{12}+J_{21})$ in order to test for consistency.

\subsection{Two views on susceptibility}\label{twoscu}

The method presented here comes from an observation about quantity $\ds \frac{\bar m_i}{\de h_j}$, which
is called $m_i$'s {\it susceptibility} with respect to external field $h_i$ in physics, or $m_i$'s {\it elasticity}
with respect to incentive $h_i$ in econometrics.

The two relevant points of view that make $\ds \frac{\bar m_i}{\de h_j}$ such an interesting quantity are those of
statistical mechanics and thermodynamics.

\subsubsection{\ref{twoscu}.1 Statistical mechanics}

For statistical mechanics $\ds \frac{\de \bar m_i}{\de h_j}$ is a quantity defined internally to the system.
The following formula clarifies this point:
From (\ref{model})
\begin{equation}\label{cij}
    \frac{\de \bar m_i}{\de h_j} = \frac{\de}{\de h_j} \Big \{ \sum_\s m_i(\s) \frac{e^{-H_N(\s)} }{Z} \Big \}=
                              \frac{N}{2} \, \big ( \o_N(m_i m_j) - \o_N(m_i) \o_N(m_j) \big ) \equiv c_{\, ij}.
\end{equation}

The quantity $\ds \frac{\de \bar m_i}{\de h_j}$, which we shall refer to as  $c_{\, ij}$ for notational convenience,
is thus simply the amount of fluctuations that we observe in quantities $m_i$: if imagine the system as a closed
box, and we imagine being inside such closed box, we can in principle measure $c_{\,ij}$ by studying the way $m_i$
vary.

\subsubsection{\ref{twoscu}.2 Thermodynamics}

The second point of view is intrinsically different: for thermodynamics $\ds \frac{\de \bar m_i}{\de h_j}$
corresponds to the response of the ``closed box'' mentioned in the last paragraph to an external influence 
given by a small change in the field $h_j$. Differently from statistical
mechanics, thermodynamics cannot provide us with this response's value a priori from observations, since
it doesn't know any details of what is going on inside the box. Thermodynamics does tell us, however, that
responses of the system to different influences , if the system is
to obey to the thermodynamic law identified by state equations (\ref{law}).

These interrelations can be made explicit by considering the partial derivatives of (\ref{law})
\begin{eqnarray*}\label{cij2}
    \frac{\de \bar m_1}{\de h_1} & = & (1-\bar m_1^2 ) \Big ( J_1 \frac{\de \bar m_1}{\de h_1} +
                                                    J_{12} \frac{\de \bar m_2}{\de h_1} +
                                                    1 \Big ),
\\
    \frac{\de \bar m_1}{\de h_2} & = & (1-\bar m_1^2 ) \Big ( J_1 \frac{\de \bar m_1}{\de h_2} +
                                                    J_{12} \frac{\de \bar m_2}{\de h_2} \Big ),
\\
    \frac{\de \bar m_2}{\de h_2} & = & (1-\bar m_2^2 ) \Big ( J_{21} \frac{\de \bar m_1}{\de h_2} +
                                                    J_{2} \frac{\de \bar m_2}{\de h_2} +
                                                    1 \Big ),
\\
    \frac{\de \bar m_2}{\de h_1} & = & (1-\bar m_2^2 ) \Big ( J_{21} \frac{\de \bar m_2}{\de h_1} +
                                                    J_{2} \frac{\de \bar m_2}{\de h_1} \Big ),
\\
\end{eqnarray*}

By relabeling $d_i=(1-\bar m_i^2)$ and using definition (\ref{cij}) we can rewrite this system of equations
as
\begin{eqnarray*}\label{cij2}
    J_1 \, c_{11} +  J_{12} \, c_{12} & = & \frac{c_{11}}{d_1} - 1 ,
\\
    J_1 \, c_{12} +  J_{12} \, c_{22} & = & \frac{c_{12}}{d_1},
\\
    J_{21} \, c_{12} +  J_{2} \, c_{22} & = & \frac{c_{22}}{d_2} - 1 ,
\\
    J_{21} \, c_{11} +  J_{2} \, c_{12} & = & \frac{c_{12}}{d_2}.
\\
\end{eqnarray*}

This is linear in the $J_{ij}$, and the former two equations are independent from the
latter two, so that we can easily solve for the $J_{ij}$ using Cramer's rule.
This together with the equations of state (\ref{law}) allows us to express all the model parameters
$J_{i,j}$ and $h_i$ as functions of the observable quantities
$\bar m_i$ and $c_{ij}$, as follows:
\begin{eqnarray*}\label{rels}
J_{12} & = & \frac{ c_{12} }{ c_{11}c_{22}-c_{12}^2 } \; = \; J_{21},
\\
J_{11} & = & \frac{\big ( \frac{c_{11}}{d_1} - 1 \big ) c_{22} - \frac{c_{12}^2}{d_1} }{c_{11} c_{22} - c_{12}^2},
\\
J_{22} & = & \frac{\big ( \frac{c_{22}}{d_2} - 1 \big ) c_{11} - \frac{c_{12}^2}{d_2} }{c_{11} c_{22} - c_{12}^2},
%
\\
\\
h_1 & = & \arctanh \, \bar m_1 - J_1 \, \bar  m_1 - J_{12} \, \bar m_2,
\\
h_2 & = & \arctanh \, \bar m_2 - J_{12} \, \bar  m_1 - J_{2} \, \bar m_2.
\\
\end{eqnarray*}

In this case we see the consistency condition $J_{12}=J_{21}$ fulfilled a priori. This tells us that, given a
set of sub-magnetizations, together with its covariance matrix, our parametrized family contains one and only one
model corresponding to it. As a consequence we can say that such model makes use of exactly the amount
of information provided into the time series of standard statistics (i.e. means and covariances) of a poll-type
database.

Estimators for $\bar m_i$ and $c_{ij}$ from the time series data are straightforward to obtain, and we
have gathered these statistics for our three case studies in Tables \ref{SM2-1}, \ref{SM2-2} and \ref{SM2-3}.
Given a time period $T$, which in our case shall correspond to a range of four consecutive years,
 we define estimators $\tilde m_i(T)$ of $\bar m_i$ and  $\tilde c_{ij}(T)$ of $\bar c_{ij}$
corresponding to it
\begin{eqnarray}\label{estimators}
\tm_i (T) &=& \frac{1}{|\, T|} \sum_{year \in T} \bar m_i^{year},
\nonumber
\\
\tilde c_{i,j} (T) &=& N_T \frac{1}{|\, T|} \sum_{year \in T} (\bar m_i^{year} - \tm_i (T))(\bar m_j^{year} - \tm_j (T)) .
\nonumber
\\
\end{eqnarray}

We must point out that in order to be well defined, such estimators should apply to a time series of 
samples which are of equal size,
since susceptibility $c_{i,j}$ has indeed an explicit size dependence. Our systems, on the other hand,
cannot be of equal size since they consist of people who chose to participate into an activity, and
the number of these people cannot be established a priori. As stated before, however, the point of view in
this thesis is that human affairs can behave following the kind of {\it quasi-static} processes familiar
to thermodynamics. Consistently with this perspective, and with some justification coming
from the considered data, we shall consider the system's population a slowly varying quantity,
and use its average of small periods of time as the quantity $N_T$ in order to define $\tilde c_{i,j} (T)$
$$
N_T= \frac{1}{|\, T|} \sum_{year \in T} N^{year}.
$$
We can thus use relations (\ref{rels}) in order to obtain estimates for the model parameters.
By considering that
\begin{eqnarray}
\a_0 &=& h_2,
\\
\a_1 &=& h_1-h_2,
\end{eqnarray}
we can compare the new estimates, presented in Tables \ref{JM2-1}, \ref{JM2-2} and \ref{JM2-3},
with those from the preceding section.

\subsection{Comments on results from the two estimation approaches}

We can now compare Tables \ref{JM2-1}, \ref{JM2-2} and \ref{JM2-3} with their counterparts
from last section, which estimated the same model for the same data coming from our three
chosen case studies, using our adaptation of Berkson's method.

Such comparison can be summarised as follows: comparing Table \ref{JM2-1}, showing parameter
estimations for the ``religious vs civil marriage'' case study, with Table \ref{J1}, we find
the estimated values to be definitely different, but we also see that they bear some interesting
similarities, especially if we consider the confidence interval provided by the least squares
method in Table \ref{J1}. Three shared features are particularly noteworthy:
    \begin{itemize}
        \item[-] The estimated values for $J_1$ and $J_2$ are similar in one aspect: in both
                  cases $J_2$ is estimated to be consistently greater than $J_1$ over the years;
        \item[-] $J_{12}$ is estimated to be very close to zero in Table \ref{JM2-1}: $J_{12}$ and
                $J_{21}$ can be considered to be statistically zero in Table \ref{J1} (which is also
                consistent with the condition $J_{12}-J_{21}=0$);
        \item[-] $\a_0$ and $\a_1$  consistently estimated with equal signs by both methods: this is
                 an essential prerequisite that any model needs to satisfy.
    \end{itemize}

The agreement is not good for the two remaining case studies, however. In the ``consensual vs non-consensual
divorce'' case study, despite estimations being consistent in the first time range (that is 2000-2003),
agreement gets worse and worse in the following two periods. As for the third case study, the two estimation
methods do not show any agreement whatsoever.

An important point to be made is the dependence of method agreement against population size. For the first
case-study, where the population is made up of over 200 thousand people the agreement between the two methods
is good. In the second case-study we have a population of roughly 40 thousand people, and we find agreement
in one of the three considered time spans. The third case-study doesn't show any agreement: the population size
here, however, is of only around 2000 people.

Finally, though the last point certainly motivates further enquiry, one should not be over-confident about
population size being the only problem. An extremely important objection comes from the fact that wherever
agreement is found, estimators $\tilde c_{\, ij}$ are found to give very high values. We must remember that
we are looking at the data through a model that assumes equilibrium: such big $\tilde c_{\, ij}$ values
correspond to large fluctuations, and these should cause an equilibrium model to be less precise and not more.

The failure of the two estimation methods to give consistent results in regimes with small fluctuations (that is
whenever $\tilde c_{\, ij}$ are small), reveal the presented study as inconclusive on an empirical level.
There are however several improvements that can be made by using the same framework established here,
the most important one concerning the handling of the data.
This thesis has as its goal to propose both a model, and a procedure allowing to establish the empirical
relevance of the model itself.
It was hence of the foremost importance to show a concrete example of such a procedure; since this was not
a professional work in statistics, however, it featured several drawbacks, some of which can be described as follows:
    \begin{itemize}
        \item[-] Though showing a remarkable temporal coherence, the time series consists of a number
                of measurements which is insufficient for any statistic to be reliable. In order to work on
                consistent groups of data, the choice was made to gather data in four-year ranges: the situation
                may be improved by considering a phenomenon having the same kind of temporal coherence, but
                for which measurements are available on a monthly basis;
        \item[-] The regional separation between ``Northern Italy'' and ``Southern Italy'' is an artificial one,
                decided for technical reasons. The quality of the statistical study could be greatly improved
                by considering a partition into groups which is directly relevant to the issue under study;
        \item[-] No use was made of the data regarding the relative sizes of the considered
                sub-populations. This, as noted before, was due to a difficulty arising from the mean-field
                assumption, which lead us to characterize the population as having equal size. This
                drawback can be amended in two ways: 1) at a fundamental level, by further considering the
                implications of having populations of different size for the model 2) by keeping the same model,
                but considering estimators for $c_{ij}$ that make use of the information coming from the subpopulation
                sizes.
    \end{itemize}

A final point to make concerns the model itself: very little is known about the structure  of the phase diagram
of a mean-field model of a multi-part system: indeed, as noted in earlier chapters, a subcase case of a two-part system
considered here
was studied in several occasions since the nineteen-fifties \cite{gorter, bidaux} until recently \cite{kulske},
and found to be highly non-trivial. As a consequence, it is to be expected that the analysis of the features
characterizing the regime that empirical data identify will need to be treated locally and numerically before any
kind of global picture arises, and it is not a priori clear whether the presence of big values for the
$c_{ij}$ might characterize and interesting regime rather than just a failure of the model. It is mainly
for this reason that much of the effort in this thesis has been directed towards the aim of establishing a way
to link the model to data, rather than to pursue further the analytic treatment of the model on its own.

\begin{table}[h]
\begin{center}
{\begin{tabular}{@{}ccccc@{}}
 & \multicolumn{4}{c}{ \textbf{4-year period} } \\
\midrule
\textbf{Statistic} & \textbf{2000-2003} & \textbf{2001-2004} & \textbf{2002-2005} & \textbf{2003-2006} \\
\midrule
$\tilde m_1$ & 0.25 & 0.20 & 0.15 & 0.12   \\
$\tilde m_2$ & 0.58 & 0.56 & 0.54 & 0.52   \\
$\tilde c_{11}$ & 1636.09 & 953.63 & 466.42 & 106.59   \\
$\tilde c_{22}$ & 346.88 & 214.58 & 122.02 & 22.30  \\
$\tilde c_{12}$ & 562.15 & 336.03 & 176.09 & 34.09   \\
\bottomrule
\end{tabular}}
\caption{Religious vs civil marriages: statistics}\label{SM2-1}
\end{center}
\end{table}

\begin{table}[h]
\begin{center}
{\begin{tabular}{@{}ccccc@{}}
 & \multicolumn{4}{c}{ \textbf{4-year period} } \\
\midrule
\textbf{Parameter} & \textbf{2000-2003} & \textbf{2001-2004} & \textbf{2002-2005} & \textbf{2003-2006} \\
\midrule
$\a_0$ & -0.21 & -0.18 & -0.15 & -0.10  \\
$\a_1$ & 0.20 & 0.17 & 0.14 & 0.08   \\
$J_1$ & 1.07 & 1.04 & 1.02 & 1.00   \\
$J_2$ & 1.51 & 1.44 & 1.39 & 1.29   \\
$J_{12}$ & 0.00 & 0.00 & 0.01 & 0.03   \\
\bottomrule
\end{tabular}}
\caption{Religious vs civil marriages: estimation of the interacting model}\label{JM2-1}
\end{center}
\end{table}

\begin{table}[h]
\begin{center}
{\begin{tabular}{@{}cccc@{}}
 & \multicolumn{3}{c}{ \textbf{4-year period} } \\
\midrule
\textbf{Statistic} & \textbf{2000-2003} & \textbf{2001-2004} & \textbf{2002-2005}  \\
\midrule
$\tilde m_1$ & 0.59 & 0.63 & 0.63   \\
$\tilde m_2$ & 0.38 & 0.45 & 0.45 \\
$\tilde c_{11}$ & 74.21 & 1.27 & 0.15 \\
$\tilde c_{22}$ & 356.27 & 1.76 & 1.68  \\
$\tilde c_{12}$ & 120.56 & -0.17 & 0.28 \\
\bottomrule
\end{tabular}}
\caption{Consensual vs non-consensual divorces: statistics}\label{SM2-2}
\end{center}
\end{table}

\begin{table}[h]
\begin{center}
{\begin{tabular}{@{}cccc@{}}
 & \multicolumn{3}{c}{ \textbf{4-year period} } \\
\midrule
\textbf{Parameter} & \textbf{2000-2003} & \textbf{2001-2004} & \textbf{2002-2005}  \\
\midrule
$\a_0$ & -0.05 & 0.23 & -0.71 \\
$\a_1$ & -0.17 & 0.01 & 5.81  \\
$J_1$ & 1.51 & 0.85 & -8.06  \\
$J_2$ & 1.16 & 0.68 & 0.39  \\
$J_{12}$ & 0.01 & -0.07 & 1.61 \\
\bottomrule
\end{tabular}}
\caption{Consensual vs non-consensual divorces: estimation of the interacting model}\label{JM2-2}
\end{center}
\end{table}

\begin{table}[h]
\begin{center}
{\begin{tabular}{@{}cccccc@{}}
 & \multicolumn{5}{c}{ \textbf{4-year period} } \\
\midrule
\textbf{Statistic} & \textbf{2000-2003} & \textbf{2001-2004} & \textbf{2002-2005} & \textbf{2003-2006} & \textbf{2004-2007} \\
\midrule
$\tilde  m_1$ & -0.29 & -0.29 & -0.29 & -0.30 & -0.28 \\
$\tilde m_2$ & -0.25 & -0.26 & -0.25 & -0.24 & -0.25 \\
$\tilde c_{11}$ & 0.94 & 0.62 & 0.30 & 0.71 & 1.95 \\
$\tilde c_{22}$ & 0.23 & 1.50 & 2.05 & 3.57 & 3.66 \\
$\tilde c_{12}$ & -0.08 & 0.00 & 0.11 & -0.50 & -0.91 \\
\bottomrule
\end{tabular}}
\caption{Suicidal tendencies: statistics}\label{SM2-3}
\end{center}
\end{table}

\begin{table}[h]
\begin{center}
{\begin{tabular}{@{}cccccc@{}}
 & \multicolumn{5}{c}{ \textbf{4-year period} } \\
\midrule
\textbf{Parameter} & \textbf{2000-2003} & \textbf{2001-2004} & \textbf{2002-2005} & \textbf{2003-2006} & \textbf{2004-2007} \\
\midrule
$\a_0$ & -1.21 & -0.16 & -0.06 & -0.13 & -0.11 \\
$\a_1$ & 0.81 & -0.29 & -0.87 & -0.37 & -0.08 \\
$J_1$ & -0.01 & -0.53 & -2.33 & -0.45 & 0.51 \\
$J_2$ & -3.40 & 0.41 & 0.57 & 0.75 & 0.76 \\
$J_{12}$ & -0.39 & 0.00 & 0.19 & -0.22 & -0.14 \\
\bottomrule
\end{tabular}}
\caption{Suicidal tendencies: estimation of the interacting model}\label{JM2-3}
\end{center}
\end{table}

%% file: ackn.tex
\chapter*{Acknowledgment}

In first place I would like to express my gratitude to my advisor Prof. Pierluigi Contucci,
for introducing me to the challenging topic of which this work gives but a glimpse,
and for carefully guiding me throughout the process of my Ph.d.
Special thanks go to Federico Gallo, who made me aware of the
subtle connection between statistical mechanics and econometrics which is central to the
thesis defended here; many
thanks are due to Cristian Giardin\`a for his fundamental role
in educating me on one of the key-points of the analytic treatment provided in this work,
and to Adriano Barra for his important contribution to the empirical case studies considered.
I would also like to thank Diego Grandi, who was available for many interesting discussions on physics
which were of great help to the completion of this task, and Stefano Ghirlanda and
Giulia Menconi with whom I was happy to collaborate in the early stage of my doctorate.